\documentclass[twocolumn]{aastex631}
\usepackage{amsmath}

\begin{document}


\title{The metallicity dependence and evolutionary times of merging binary black holes: Combined constraints from individual gravitational-wave detections and the stochastic background}


\author[0000-0002-9296-8603]{Kevin Turbang}
\affiliation{Theoretische Natuurkunde, Vrije Universiteit Brussel, Pleinlaan 2, B-1050 Brussels, Belgium}
\affiliation{Universiteit Antwerpen, Prinsstraat 13, B-2000 Antwerpen, Belgium}
\email{kevin.turbang@vub.be}

\author[0000-0002-2254-010X]{Max Lalleman}
\affiliation{Universiteit Antwerpen, Prinsstraat 13, B-2000 Antwerpen, Belgium}

\author[0000-0001-9892-177X]{Thomas A. Callister}
\affiliation{Kavli Institute for Cosmological Physics, The University of Chicago, 5640 S. Ellis Ave., Chicago, IL 60615, USA}

\author[0000-0003-4180-8199]{Nick van Remortel}
\affiliation{Universiteit Antwerpen, Prinsstraat 13, B-2000 Antwerpen, Belgium}
\begin{abstract}
The advent of gravitational-wave astronomy is now allowing for the study of compact binary merger demographics throughout the Universe. This information can be leveraged as tools for understanding massive stars, their environments, and their evolution. One active question is the nature of compact binary formation: the environmental and chemical conditions required for black hole birth and the time delays experienced by binaries before they merge. Gravitational-wave events detected today, however, primarily occur at low or moderate redshifts due to current interferometer sensitivity, therefore limiting our ability to probe the high redshift behavior of these quantities. In this work, we circumvent this limitation by using an additional source of information: observational limits on the gravitational-wave background from unresolved binaries in the distant Universe. Using current gravitational-wave data from the first three observing runs of LIGO-Virgo-KAGRA, we combine catalogs of directly detected binaries and limits on the stochastic background to constrain the time-delay distribution and metallicity dependence of binary black hole evolution. Looking to the future, we also explore how these constraints will be improved at the Advanced LIGO A+ sensitivity. We conclude that, although binary black hole formation cannot be strongly constrained with today's data, the future detection (or a non-detection) of the gravitational-wave background with Advanced LIGO A+ will carry strong implications for the evolution of binary black holes.
\end{abstract}


\section{Introduction} \label{s:intro}
Since the first detection of gravitational waves by the LIGO-Virgo collaborations in 2015 \citep{PhysRevD.88.043007, LIGOScientific:2014pky, VIRGO:2014yos, PhysRevLett.116.061102}, the number of detected compact binary coalescences has been steadily increasing, amounting to around 90 detections in the third observing run \citep{PhysRevX.13.011048}. These individual detections offer new insights into the demographics of compact binary mergers, including their mass and spin distributions. At the same time, growing gravitational-wave catalogs offer an opportunity to study the evolution of the binary merger rate over cosmic time. The redshift-dependent merger rate is dictated by a confluence of environmental conditions governing compact binary formation and evolutionary processes describing their subsequent evolution. It therefore offers a window through which we can learn about the births of compact binaries and the lives of their progenitors. 

Compact binaries are expected to preferentially arise from massive stars in low metallicity environments \citep{Fryer_2012, Belczynski_2016,Chruslinska_2018,  Mapelli_2019, Santoliquido_2020,Liotine_2023}. Both the underlying metallicity-specific star formation rate and the metallicity dependence of compact object formation, however, remain poorly understood \citep{10.1093/mnras/stz2057, 2021MNRAS.508.5028B,Broekgaarden_2022, chruslinska2022chemical, chruslinska2023trading, van_Son_2023}. Therefore, gravitational-wave observations might provide one of the best observational routes to constraining or measuring the metallicity specific star formation rate and compact object formation efficiency. Furthermore, the time-delay distribution between progenitor formation and eventual binary merger is not known and, if measured, it could provide hints as to what are the dominant formation channels and environments of binary black holes. Binaries in the field, for example, exhibit longer/shorter evolutionary time delays if they undergo stable/unstable (e.g. common envelope) mass transfer, respectively \citep{van_Son_2022}. Binaries assembled dynamically in clusters might experience a different range of time delays altogether, with further distinctions depending on whether these dynamical systems merge inside the cluster or are jettisoned and merge outside the cluster \citep{Samsing_2018}. 

Current gravitational-wave detectors, however, are limited in their ability to probe high-redshift binary black hole merger events. For example, current catalogs provide meaningful constraints on the merger rate only to $z\lesssim1$ \citep{PhysRevX.13.011048,callister2023parameterfree}, although the formation rate can be probed out to $z\sim 4$ for population synthesis-motivated assumptions about the time-delay distribution \citep{fishbach2023ligovirgokagras}. Even with the future Advanced LIGO A+ sensitivity \citep{AplusSensitivity}, the redshifts that will be probed by individual compact binary detections will likely remain below $z \sim 3$, thus suggesting that the high-redshift investigation of the binary black hole merger rate will remain challenging, at least until the era of next-generation instruments like the Einstein Telescope and Cosmic Explorer \citep{Vitale_2019,evans2021horizon,borhanian2022listening}.

Direct compact binary detections are not our only source of information, however. In addition to those individual binaries detected by the LIGO-Virgo-KAGRA collaborations, a gravitational-wave background from unresolvable compact binaries is expected \citep{Romano_2017, Christensen_2018, VANREMORTEL2023104003}. This gravitational-wave background can be detected in the form of excess correlated noise shared between distant gravitational-wave detectors. Even though a gravitational-wave background remains to be detected by the LIGO-Virgo-KAGRA collaborations, current upper limits on the gravitational-wave background can already allow us to constrain the merger rate at higher redshifts than individual binary black hole merger detections alone \citep{Abbott_2021, Abbott_2021_anisotropic}.

Many studies have previously used catalogs of gravitational-wave detections to study the formation history of compact binaries \citep{Fishbach_2018, Vitale_2019, Fishbach_2021, Mukherjee_2022,Riley_2023}. In this paper, we will study the evolutionary history of compact binaries by combining direct detections with constraints on the gravitational-wave background. The approach of combining individual binary black hole merger events with upper limits on the gravitational-wave background was first explored in \cite{Callister_2020}, where a phenomenological broken power-law model was adopted for the binary black hole merger rate. In this study, however, we will seek to go a step further, moving past the merger rate itself and constraining the metallicity dependence and evolutionary time delays associated with compact binary formation and evolution.

We find that current data can constrain the slope of the binary time-delay distribution, but does not contain information regarding the metallicity dependence of compact binary formation. Moreover, current limits on the stochastic gravitational-wave background are not yet strong enough to noticeably impact these results. However, we show that the sensitivity of future A+ instruments will dramatically change this picture. At A+ sensitivities, both the detection or non-detection of the gravitational-wave background will provide novel constraints on the parameters governing these distributions, offering a unique probe of the environment in which the binary formed and its evolution.

This work is structured as follows. In Sec.~\ref{s:BBH}, we introduce our model for the binary black hole merger rate, starting from a metallicity-dependent star formation rate and an evolutionary time-delay distribution. In parallel, we demonstrate how differing assumptions regarding binary black hole formation impact the expected amplitude of the gravitational-wave background. 
The methodology for our joint individual binary black hole and stochastic background analysis method is outlined in Sec.~\ref{s:Analysis}. Our results using data from the first three LIGO-Virgo-KAGRA observing runs are discussed in Sec.~\ref{ss:O3Results}. Finally, we explore the potential of the analysis method with future constraints by considering the Advanced LIGO A+ sensitivity in Sec.~\ref{ss:O5Results}.

\section{Binary black hole merger rate}
\label{s:BBH}

Whereas past work \citep{Callister_2020} focused on characterizing the resulting merger rate of black holes, in this paper we will seek to work further backwards, studying the underlying formation efficiency and time-delay distribution of binary black holes. Standard models for the merger history of compact binaries rely on three ingredients: a metallicity specific star formation rate, an efficiency with which stellar progenitors yield merging compact objects, and a distribution of evolutionary time delays.

Throughout this work, we will consider the star formation rate as given by \cite{MDSFR}.
They obtain a fit to the global star formation rate per unit volume of the form: 
\begin{equation}
\label{eq:SFR-Madau}
    R_*(z)\propto \frac{(1+z)^{2.7}}{1+\left((1+z)/2.9\right)^{5.6}}.
\end{equation}
In App. \ref{app:resultsVangioni}, we provide results for an alternative star formation rate given by \cite{10.1093/mnras/stu2600} to illustrate that the main conclusions of this work can still be drawn if one assumes a different star formation rate.

\begin{figure*}
    \includegraphics[scale=0.48]{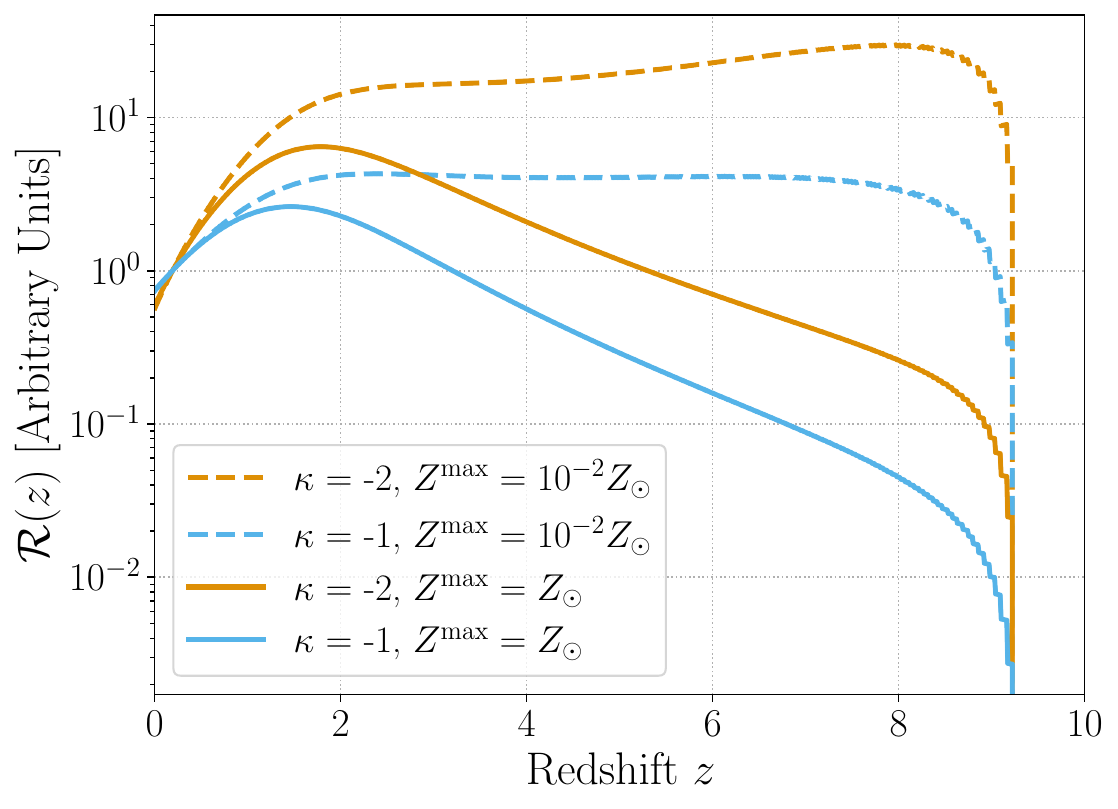}
    \includegraphics[scale=0.48]{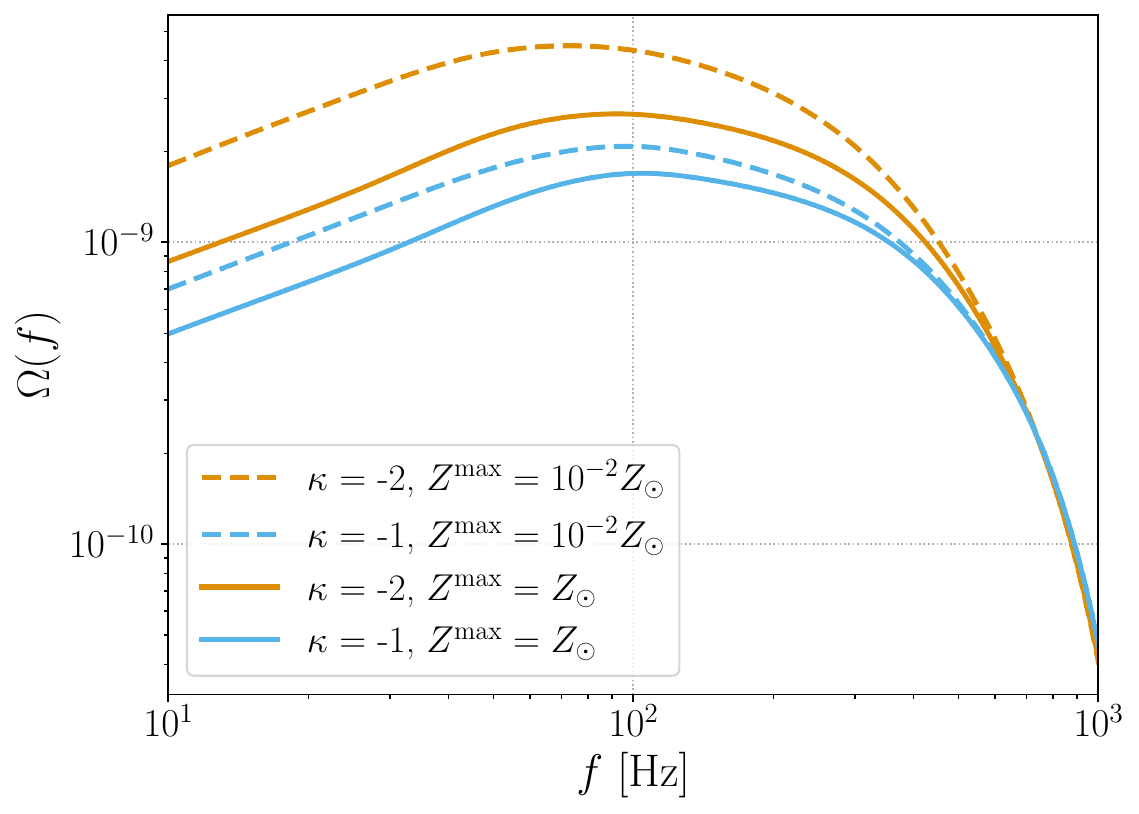}
    \caption{\textbf{Left:} Examples of the binary black hole merger rate $\mathcal{R}(z)$ for different values of the slope of the time-delay distribution, $\kappa$, and different values of the maximum metallicity below which black holes are formed, $Z^{\rm max}$. All curves are arbitrarily normalized to $\mathcal{R}(z=0.2)=1$. The value of the minimum time-delay parameter, $t_d^{\rm min}$, is set to 0.05 Gyrs. Steeper time-delay distributions result in a larger merger rate at large redshifts, whereas larger metallicity thresholds result in a lower merger rate at large redshifts. \textbf{Right:} Example of $\Omega(f)$ spectrum for different values of the same parameters $\kappa$ and $Z^{\rm max}$. The value of all other parameters that enter in the mass distribution as given by Eqs. \eqref{eq:p(m1)} - \eqref{eq:p_q} is fixed to arbitrary values: $m_{\rm low}=10M_\odot$, $m_{\rm high}=80M_\odot$, $\alpha=-2$, $\mu_m=20M_\odot$, $\sigma_m=3M_\odot$, $f_p=0.01$, $\beta_q=2$, $\mathcal{R}_{\rm ref}=1\,M_\odot^{-1}\,{\rm Gpc^{-3}}\,{\rm yr^{-1}}$, $\delta m_{\rm low}=1 M_\odot$, $\delta m_{\rm high}=3 M_\odot$, and $t_d^{\rm min}=0.05$ Gyrs. For more details on the time-delay parameter $t_d^{\rm min}$ and $\mathcal{R}_{\rm ref}$, we refer the reader to Eqs. \eqref{eq:p(td)} and \eqref{eq:R20}, respectively. The behavior of the $\Omega(f)$ spectrum as the time-delay parameters are varied can directly be related to the behavior of the merger rate on the left-hand side, as a larger integrated merger rate results in a larger net $\Omega(f)$ spectrum.}
    \label{fig:MergerRatePlot}
\end{figure*}

It is expected that merging binary black holes preferentially form in low metallicity environments, with efficiencies falling steeply at metallicities $Z \gtrsim 0.1~Z_\odot$ \citep{Belczynski_2016, Chruslinska_2018,Klencki_2018,Mapelli_2019,2019MNRAS.490.3740N,Santoliquido_2020,2021MNRAS.502.4877S,Broekgaarden_2022}.
Motivated by this, we will assume that the birth rate of binary black holes is proportional to the product $R_*(z) F(Z^{\rm max},z)$, where $F(Z^{\rm max},z)$ is the fraction of star formation occurring at redshift $z$ below metallicity $Z^{\rm max}$. An analytic prescription for the fraction of star formation is given by \cite{2006ApJ...638L..63L} as
\begin{equation}
\label{eq:metallicity_equation}
    F(Z^{\rm max},z)=\frac{1}{1.122}\hat{\Gamma}\left(0.84,\left(\frac{Z^{\rm max}}{Z_\odot}\right)^2 10^{0.3z}\right),
\end{equation}
where $\hat{\Gamma}$ is the incomplete gamma function and $Z_\odot$ is the solar metallicity. The merger rate as a function of redshift is then obtained by performing a convolution of the metallicity-dependent star formation rate and a distribution of time delays $t_d$ between binary formation and merger. We model this time-delay distribution as a power law,
\begin{equation}
\label{eq:p(td)}
    p\left(t_d|\kappa, t_d^{\rm min} \right) \propto 
    \begin{cases}
    (t_d)^{\kappa}  &\left(t_d^{\min } \leq t_d \leq t_d^{\max }\right) \\ 
    0 &(\text {else})
    \end{cases},
\end{equation}
where $t_d^{\rm max}$ is fixed to the age of the Universe, i.e., 13.5 Gyrs, and $t_d^{\rm min}$ is a parameter that can be inferred from the data. 
The merger rate as a function of redshift is now given by the convolution of the time-delay distribution with the metallicity-dependent star formation rate:
\begin{equation}
\label{eq:mergerRateBBH}
\mathcal{R}(z) = \int dt_d~p(t_d) R_*\left(z_f(z, t_d)\right)F\left(Z\leq Z^{\rm max}, z_f(z,t_d)\right),
\end{equation}
where the binary formation redshifts $z_f = z_f(z,t_d)$ are regarded as a function of merger redshift and time delay. 

Our prescription for the binary black hole merger rate left three parameters undefined -- the slope $\kappa$ of the time-delay distribution, the minimum time delay $t_d^{\rm min}$, and the maximum metallicity $Z^{\rm max}$ amenable to compact binary formation. Our goal in this work will be to \textit{infer} these parameters from gravitational-wave observations. The measurement of these parameters could shed light on the different binary formation channels, as each of these predicts different values of the spectral index $\kappa$. For example, the value of $\kappa$ is expected to be -1 in the classical field formation scenario \citep{2012ApJ...759...52D,Dominik_2013, fishbach2023ligovirgokagras}. 

An example of the merger rate $\mathcal{R}(z)$ is given in the left-hand side of Figure~\ref{fig:MergerRatePlot}. We vary the slope of the time-delay distribution $\kappa$, as well as the maximum metallicity below which black hole formation occurs $Z^{\rm max}$, but fix the value of the minimum time-delay parameter $t_d^{\rm min}$ to an arbitrary value of 0.05 Gyrs. Each curve is arbitrarily normalized to $\mathcal{R}(z=0.2)=1$. A more negative slope $\kappa$ corresponds to more events that merged at early evolutionary times. Given a fixed \textit{observed} merger rate in the local Universe, more negative $\kappa$ correspondingly boosts the merger rate at large redshifts, as illustrated in the left-hand side of Figure~\ref{fig:MergerRatePlot}. Similarly, lower metallicity thresholds allow for a larger merger rate at earlier times, i.e., at large redshifts, when the metallicity of the environment in which the binaries formed was still low. Note that, although the effect of varying the minimum time-delay parameter is not displayed, decreasing its value would result in an increased merger rate at high redshifts, as mergers now occur at earlier times, and therefore, larger redshifts.

As we vary parameters, the behavior of the black hole merger history in the local Universe, where binary black hole mergers are currently observed, remains practically unaltered \citep{PhysRevX.13.011048}. In contrast, the most dramatic differences occur at large redshifts. Although these redshifts are too distant to directly observe, these different models yield very different predictions regarding the stochastic gravitational-wave background.


The gravitational-wave background is expressed in terms of the dimensionless energy density fraction \citep{Romano_2017}:
\begin{equation}
\Omega(f)=\frac{1}{\rho_{c}} \frac{d \rho_{\mathrm{GW}}}{d \ln f},
\end{equation}
where $\rho_c=3H_0^2c^2/(8\pi G)$ is the critical energy density of the Universe, $G$ is Newton's gravitational constant, $c$ is the speed of light, and the Hubble constant is given by $H_0=67.9$ km s$^{-1}$ Mpc$^{-1}$ \citep{2016A&A...594A..13P}. More specifically, for the gravitational-wave background coming from unresolved compact binaries throughout the Universe, the dimensionless energy density takes the form \citep{phinney2001practical,Callister_2020}:
\begin{equation}
\label{eq:OmegaBBH}
    \Omega(f)=\frac{f}{\rho_{c}} \int_{0}^{z_{\max }} d z\, \frac{\mathcal{R}(z)\Big\langle\dfrac{d E_{s}}{d f_{s}}\Big\rangle\Big|_{f(1+z)}}{(1+z) H(z)},
\end{equation}
where the Hubble rate at redshift $z$ is given by $H(z)=H_0\sqrt{\Omega_M(1+z)^3+\Omega_\Lambda}$, neglecting the contribution from the radiation density, and $\mathcal{R}(z)$ denotes the binary black hole merger rate. The matter and dark energy densities are taken to be $\Omega_M= 0.3$ and $\Omega_\Lambda=0.7$, respectively \citep{2016A&A...594A..13P}. We impose a redshift cut-off $z_{\rm max}=10$, as few binary black holes are expected beyond this redshift, provided the binary black holes are stellar progenitors. The quantity $\langle dE_s/df_s\rangle$ is the gravitational-wave energy spectrum of a single binary, averaged across the binary black hole population \citep{PhysRevD.77.104017}. If $\lambda$ denotes the intrinsic properties of a given binary (masses, spins, etc.) and $p(\lambda)$ is the distribution of these parameters across the binary black hole population, then
\begin{equation}
\label{eq:AverageEnergy}
    \left\langle\frac{d E_{s}}{d f_{s}}\right\rangle=\int d \lambda p(\lambda) \frac{d E_{s}}{d f_{s}}(\lambda),
\end{equation}
where this quantity is evaluated at the source-frame frequency $f_s=f(1+z)$ in Eq.~\eqref{eq:OmegaBBH}. More details about the assumed mass and spin distributions will be given below.

In the right-hand side of Figure \ref{fig:MergerRatePlot}, we show an example of the computed $\Omega(f)$ spectrum for different values of the time-delay distribution parameters, as introduced in Eqs.~\eqref{eq:p(td)} and \eqref{eq:mergerRateBBH}. The variation in the $\Omega(f)$ spectrum when varying the time-delay distribution and metallicity parameters can be directly linked to the behavior of the merger rate $\mathcal{R}(z)$ in the left-hand side of Figure~\ref{fig:MergerRatePlot}. Indeed, increasing the integrated merger rate correspondingly increases the $\Omega(f)$. Note that the largest background comes from a steep time-delay distribution and a very low maximum metallicity, while the smallest background comes from allowing long time delays and a high maximum metallicity. For a more detailed explanation of the computation of $\Omega(f)$, we refer the reader to App.~\ref{app:computationOmega}. Furthermore, we note that the example given here is largely illustrative. More detailed investigations have recently been performed by \cite{lehoucq2023astrophysical}.

\section{Analysis method}
\label{s:Analysis}

Our goal is to learn about the evolutionary history of binary black holes by synthesizing all available gravitational-wave information, combining both direct detections of black hole mergers with upper limits on the integrated gravitational-wave background, based on work by \cite{Callister_2020}. The inputs to our analysis are a set of $N_{\rm obs}$ direct binary black hole detections, with data $\{d\}^{N_{\rm obs}}_{i=1}$, along with cross-correlation measurements $\hat{C}(f)$ of the gravitational-wave background (to be explained further below). Let $\Lambda$ signify the set of hyperparameters describing the binary black hole population, e.g., the slope of the time-delay distribution, $\kappa$, introduced in Eq. \eqref{eq:p(td)}. We assume that the joint likelihood of these measurements can be factorized as
\begin{equation}
\label{eq:JointLikelihood}
    p\bigl(\left\{d_{i}\right\},\hat{C}|\Lambda\bigr)=p_{\text {BBH}}\left(\left\{d_{i}\right\}|\Lambda\right) p_{\text{GWB}}(\hat{C}|\Lambda),
\end{equation}
where $p_{\rm BBH}$ and $p_{\rm GWB}$ denote the likelihoods for the individual binary black hole detections and the gravitational-wave background, respectively.\footnote{This factorization does not strictly hold, as the same stretches of data contribute to both the direct compact binary detections and the gravitational-wave background measurements. However, the contributions of direct detections to gravitational-wave background measurements are currently negligible~\citep{Abbott_2021}, making the factorization a good approximation.}

The individual binary black hole likelihood takes the form \citep{10.1063/1.1835214, PhysRevD.98.083017,10.1093/mnras/stz896}:
\begin{equation}
\label{eq:likelihood1BBH}
  p_{\mathrm{BBH}}\left(\left\{d_{i}\right\}|\Lambda\right)\propto e^{-N_{\rm exp}\left(\Lambda\right)} \prod_{i=1}^{N_{\mathrm{obs}}}\int p\left(d_{i}|\lambda\right) \frac{dN}{d\lambda}(\Lambda) d \lambda,
\end{equation}
where, as in Sec.~\ref{s:BBH}, $\lambda$ denotes the parameters (redshift, masses, spins, etc.), and $p(d_i|\lambda)$ denotes the likelihood of a gravitational-wave event $d_i$ given parameters $\lambda$. In the expression above, we also note the presence of the differential mass-redshift distribution for binary black holes, which is given by:
\begin{equation}
\label{eq:diffdistribution}
    \frac{dN}{dzdm_1dm_2} = Np(z,m_1,m_2),
\end{equation}
where $N$ denotes the total number of binary black hole mergers integrated across all redshifts and masses.
Assuming that the component mass distributions do not vary with redshift,\footnote{
Although intrinsic evolution of the compact binary mass distribution with redshift is a generic prediction~\citep{van_Son_2022,2024arXiv240212444Y}, no such evolution is detected with current data~\citep{2021ApJ...912...98F,PhysRevX.13.011048,van_Son_2022}.
} this allows us to decompose the mass-redshift distribution and adopt the following parameterization:
\begin{equation}
    \label{eq:paramMergerRate}
    \frac{dR}{dm_1dm_2}=\mathcal{R}_{\rm ref}\frac{\mathcal{R}(z)}{\mathcal{R}(0.2)}\frac{\phi(m_1)}{\phi(20M_\odot)}p(m_2),
\end{equation}
where
\begin{equation}
\label{eq:R20}
    \mathcal{R}_{\rm ref}=\left.\frac{dR}{dm_1}\right|_{z=0.2,~m_1=20M_\odot}
\end{equation}
is the differential source-frame merger rate at redshift $z=0.2$ and primary component mass $m_1=20M_\odot$, and $\phi(m_1)$ and $p(m_2)$ characterize the primary and secondary mass distributions, respectively.
The source-frame merger rate in Eq.~\eqref{eq:paramMergerRate} can be related to the detector-frame rate in Eq. \eqref{eq:diffdistribution} by
\begin{equation}
    \frac{dN}{dzdm_1dm_2} = \frac{dV_c}{dz}\frac{1}{1+z}\frac{dR}{dm_1dm_2},
\end{equation}
where the factor $(1+z)^{-1}$ transforms from detector-frame to source-frame time, and $\frac{dV_c}{dz}$ denotes the comoving volume per unit redshift.

Following results from the LIGO-Virgo-KAGRA GWTC-3 catalog \citep{PhysRevX.13.011048}, we model the primary mass distribution as a mixture between a power-law and an additional Gaussian excess:
\begin{multline}
\label{eq:p(m1)}
    p(m_1)=\frac{f_p}{\sqrt{2\pi\sigma_m^2}}\exp\left(-\frac{(m_1-\mu_m)^2}{2\sigma_m^2}\right)\\+(1-f_p)\left(\frac{1+\alpha}{(100M_\odot)^{1+\alpha}-(2M_\odot)^{1+\alpha}}\right)m_1^\alpha,
\end{multline}
where $\alpha$ is the spectral index of the power-law, the mean $\mu_m$ and variance $\sigma_m^2$ characterize the Gaussian peak, and $f_p$ denotes the relative contribution of the Gaussian peak and the power-law \citep{Talbot_2018, PhysRevX.11.021053}. A smoothing function is applied, such that the mass distribution is suppressed for $m_1<m_{\rm low}$ and $m_1>m_{\rm high}$:
\begin{equation}
\label{eq:phi_m}
\hspace{-.7cm}
    \phi\left(m_1\right)=\left\{\begin{array}{ll}
p\left(m_1\right) \exp \left(\frac{-\left(m_1-m_{\text {low }}\right)^2}{2 \delta m_{\text {low }}^2}\right) & \left(m_1<m_{\text {low }}\right) \\
p\left(m_1\right) \exp \left(\frac{-\left(m_1-m_{\text {high }}\right)^2}{2 \delta m_{\text {high }}^2}\right) & \left(m_1 > m_{\text {high }}\right) \\
p\left(m_1\right) & \left(\mathrm{else}\right)
\end{array}\right.
\end{equation}
where $\delta m_{\rm low}$ and $\delta m_{\rm high}$ denote the scale of smoothing. 
The secondary mass $m_2$ is assumed to follow a power-law distribution, while ensuring $m_2<m_1$:
\begin{equation}
\label{eq:p_q}
    p(m_2|m_1) = \left(\frac{1+\beta_q}{m_1^{1+\beta_q}-(2M_\odot)^{1+\beta_q}}\right)m_2^{\beta_q},
\end{equation}
where $\beta_q$ denotes the spectral index of the distribution, and $p(m_2|m_1)$ is set to zero below $2M_\odot$.
Although not explicitly mentioned throughout this section, we additionally measure the distributions of component spin magnitudes and spin-orbit tilt angles; see App. \ref{app:models}.


In Eq.~\eqref{eq:likelihood1BBH}, the total number of gravitational-wave events, i.e., detected and undetected, is denoted by $N(\Lambda)$, whereas the number of observed events is represented by $N_{\rm obs}$. The expected number of observations $N_{\rm exp}$ is given by 
\begin{equation}
    N_{\rm exp}(\Lambda)=\int d \lambda P_{\text{det}}(\lambda) \frac{dN}{d\lambda}(\lambda|\Lambda),
\end{equation}
where $P_{\text{det}}(\lambda)$ is the probability of successfully detecting a binary black hole merger described by parameters $\lambda$. In practice, $N_{\rm exp}(\Lambda)$ can be approximated through a Monte Carlo average over artificial signals injected into detector data:
\begin{equation}
    \label{eq:MC_selection_effects}
    N_{\rm exp}(\Lambda) = \frac{1}{N_{\rm inj}}\sum_i^{N_{\rm found}}\frac{dN/d\lambda_i}{p_{\rm inj}(\lambda_i)},
\end{equation}
where the sum runs over the number of above detection threshold injections, $N_{\rm found}$, drawn from a reference distribution $p_{\rm inj}(\lambda_i)$ \citep{Callister_2022}. More details about the set of injections used in this analysis can be found in App. \ref{app:Data}.

The remaining necessary element is the integration over individual binary black hole likelihoods $p(d_i|\lambda)$ in Eq.~\eqref{eq:likelihood1BBH}.
This can be estimated using Monte Carlo integration over posterior samples drawn from a reference prior \citep{10.1063/1.1835214, PhysRevD.98.083017,10.1093/mnras/stz896}:
\begin{multline}\label{eq:BBHLikelihood}
    p_{\text {BBH}}\left(\left\{d_{i}\right\}|\Lambda\right)\propto e^{-N_{\rm exp}\left(\Lambda\right)} \prod_{i=1}^{N_{\text {obs}}}\left\langle\frac{dN/d\lambda_i}{p_{\text {pe}}\left(\lambda_{i}\right)}\right\rangle_{\text {samples}},
\end{multline}
where $p_{\rm pe}(\lambda_i)$ denotes the prior used during the parameter estimation to infer the distribution of the parameters describing the gravitational-wave event, and the average is taken over the posterior samples of each event. 

Our joint analysis also incorporates the contribution from the gravitational-wave background. Gravitational-wave background searches, such as the one performed by the LIGO-Virgo-KAGRA collaborations \citep{Abbott_2021}, aim to measure the dimensionless energy density $\Omega(f)$ introduced above. To this end, an optimal cross-correlation estimator is defined \citep{Allen_1999}:
\begin{equation}
\label{Eq: C_hat}
\hat{C}(f)=\frac{2}{T} \frac{10 \pi^{2}}{3 H_{0}^{2}} \frac{f^{3}}{\gamma_{12}(f)} \tilde{s}_{1}(f) \tilde{s}_{2}^{*}(f),
\end{equation}
where $T$ is the observation time, and $\tilde{s}_1(f)$ and $\tilde{s}_2(f)$ denote the Fourier transformed data in both detectors. The overlap reduction function $\gamma_{12}(f)$ encodes the baseline geometry of the detector pair \citep{PhysRevD.46.5250,PhysRevD.48.2389}. The normalization of this estimator is such that $\langle\hat{C}(f)\rangle= \Omega(f)$. An estimator of the variance is given by $\left<\hat{C}(f)\hat{C}(f^\prime)\right>=\delta(f-f^\prime)\sigma^2(f)$, with
\begin{equation}
\label{Eq: sigma_hat}
\sigma^{2}(f)\approx\frac{1}{2T\Delta f}\left(\frac{10 \pi^{2}}{3 H_{0}^{2}}\right)^{2} \frac{f^{6}}{\gamma_{12}^2(f)} P_{1}(f) P_{2}(f),
\end{equation}
under the assumption of a small signal-to-noise ratio, where $P_1(f)$ and $P_2(f)$ denote the one-sided power spectra in both detectors, and $\Delta f$ is the frequency resolution.

This cross-correlation estimator then enters in the likelihood $p_{\rm GWB}$ in Eq.~\eqref{eq:JointLikelihood}, which is well-approximated by a Gaussian distribution \citep{Abbott_2021}:
\begin{equation}
p_{\text {GWB}}\left(\hat{C}|\Lambda\right)\propto \exp \left[-\frac{1}{2}\sum_k\left(\frac{\hat{C}(f_k)- \Omega\left(f_k,\Lambda\right)}{\sigma(f_k)}\right)^2\right],
\end{equation}
where the sum runs over discrete frequency bins $f_k$ and $\Omega(f)$ is given by Eq.~\eqref{eq:OmegaBBH}. For additional information, we refer the reader to \cite{Abbott_2021}. 

All together, we infer 16 hyperparameters: the nine parameters governing the primary and secondary mass distributions in Eqs. \eqref{eq:phi_m} and \eqref{eq:p_q}, the three spin parameters detailed in App. \ref{app:models}, the three parameters $\{Z^{\rm max},~\kappa,~t_d^{\rm min}\}$ characterizing the formation and evolution of binary black hole systems, as well as the reference merger rate amplitude $\mathcal{R}_{\rm ref}$ defined in Eq. \eqref{eq:R20}. The priors used for the parameter estimation are summarized in Table~\ref{tab:priors} in App. \ref{app:models}.

\section{Results}
\label{s:Results}
In this section, we discuss the results of our analyses. Sec. \ref{ss:O3Results} presents the results using currently available LIGO-Virgo data, while Sec. \ref{ss:O5Results} shows results that will be available in future observing runs with the improved A+ sensitivity of the Advanced LIGO detectors.
\subsection{Constraints from LIGO-Virgo O1, O2, and O3 data}
\label{ss:O3Results}
\begin{figure*}
    \centering
    \includegraphics[width = .9\textwidth]{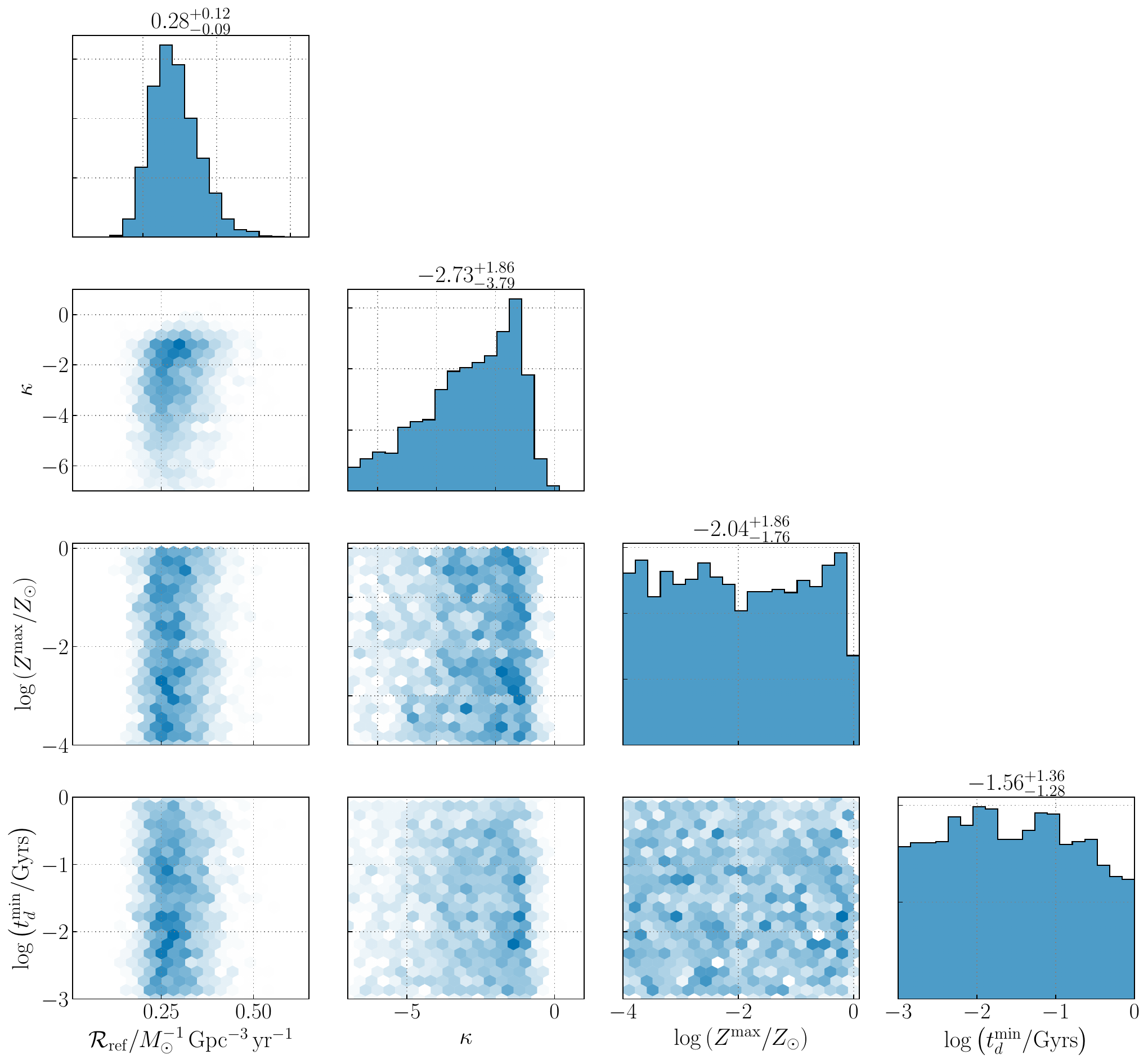}
    \caption{Posteriors on the parameters governing binary black hole birth and evolution using both direct detections and stochastic background constraints on the first three LIGO-Virgo observing runs.
    For clarity, the only posteriors that are shown are for the reference merger rate amplitude $\mathcal{R}_{\rm ref}$, the slope of the time-delay distribution $\kappa$, the metallicity threshold $Z^{\rm max}$, and the minimum time delay $t_d^{\rm min}$, although we simultaneously inferred the binary black hole mass distribution as in Sec.~\ref{s:Analysis}. The reference merger rate amplitude $\mathcal{R}_{\rm ref}$ is well constrained due to the currently observed individual binary black hole merger events happening at low redshifts. Although the metallicity threshold $Z^{\rm max}$ and the minimum time-delay parameter $t_d^{\rm min}$ cannot be constrained with data from the first three observing runs, the slope of the time-delay distribution $\kappa$ is constrained to negative values. Positive values of this parameter would give rise to a different merger rate that decreases with redshift, contradicting current observations.}
    \label{fig:posteriorO3}
\end{figure*}

We now apply the methodology outlined in the previous sections to place constraints on the time-delay parameters giving rise to the binary black hole population, using data from LIGO-Virgo's first three observing runs. 
More details about the exact data used are given in App. \ref{app:Data}.

\begin{figure*}
    \includegraphics[scale = .47]{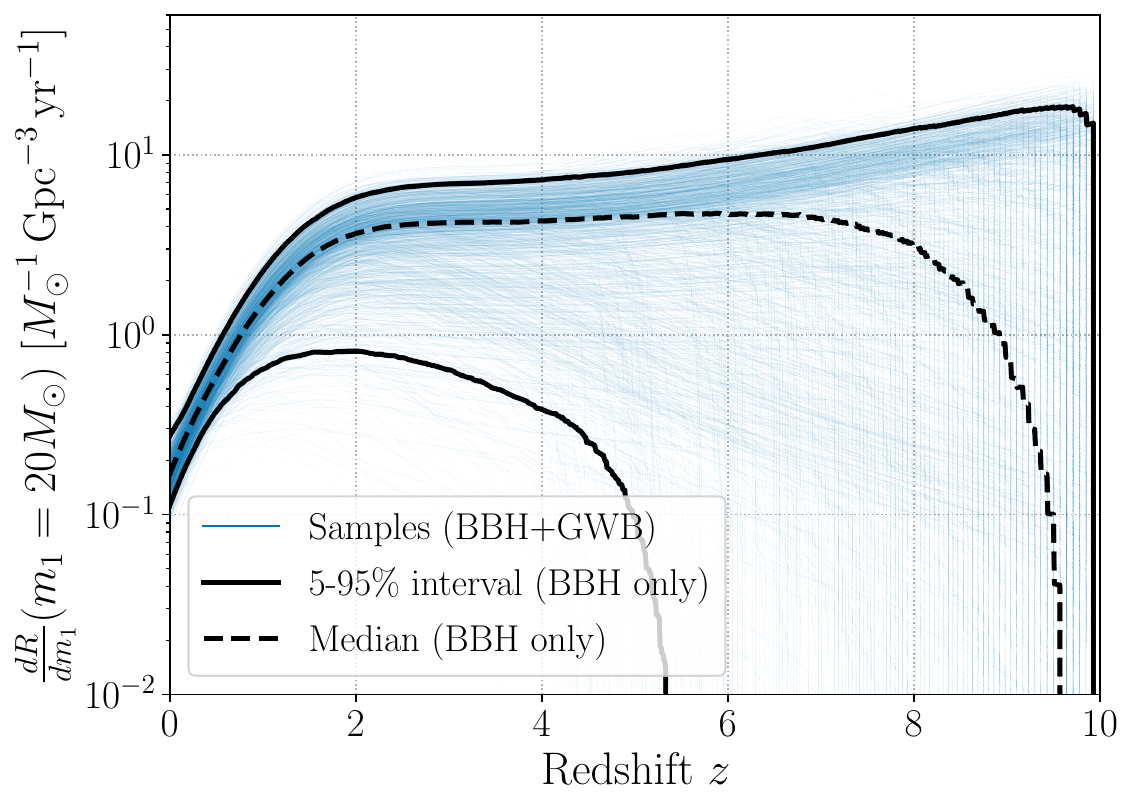}
    \includegraphics[scale = .47]{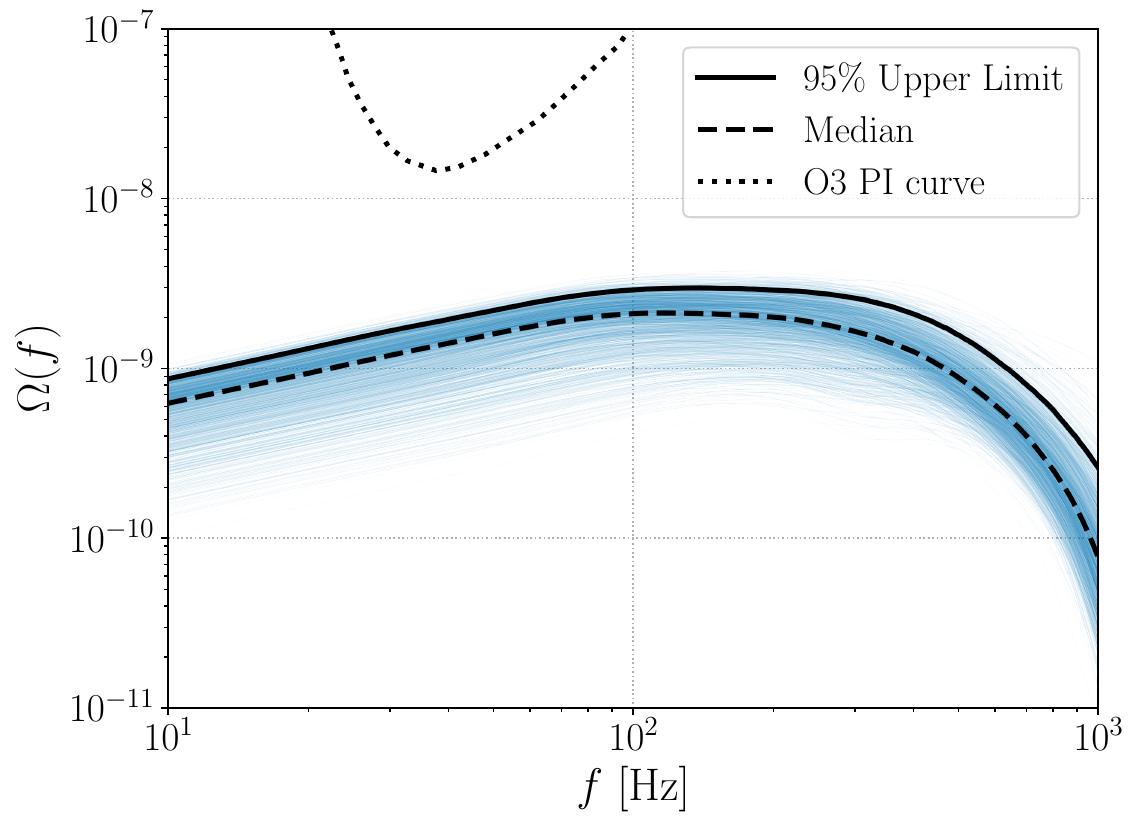}
    \caption{\textbf{Left}: Posterior samples on the merger rate $\frac{dR}{dm_1}(m_1=20M_\odot)$ for the joint analysis using both direct detections and stochastic background constraints.
    For reference, the 5\% and 95\% credible limits are shown in full black lines for the individual binary black hole run, and the dashed black line shows the median value for the binary black hole only run. The results of the individual binary black hole run and the joint analysis using both direct detections and stochastic background constraints are comparable, illustrating that the gravitational-wave background upper limits are not informative at current detector sensitivity. Furthermore, we note that the merger rate is well-constrained at low redshifts because of the individual binary black hole mergers detected at those redshifts. \textbf{Right}: Posterior on $\Omega(f)$ using the posterior samples from the joint analysis using both direct detections and stochastic background constraints on the first three LIGO-Virgo-KAGRA observing runs. The median is denoted by the dashed black line, and the 95\% credible upper limit by the full black line. The $2\sigma$ O3 PI curve from \cite{Abbott_2021} is shown (dotted black) as an indication of the gravitational-wave background detection threshold after the first three LIGO-Virgo observing runs. All posterior samples fall below the O3 PI curve, illustrating that we do not expect the gravitational-wave background contribution to be informative at current detector sensitivity.}
\label{fig:mergerRateO3samples}
\end{figure*}

In Figure \ref{fig:posteriorO3}, we show the obtained posteriors on the time-delay parameters $\{\kappa,~Z^{\rm max}, t_d^{\min}\}$ from Eqs.~\eqref{eq:p(td)} and \eqref{eq:mergerRateBBH}, as well as the merger rate amplitude $\mathcal R_{\rm ref}$ at $z=0.2$ and $m_1=20M_\odot$, as defined by Eq.~\eqref{eq:R20}. We simultaneously infer all the hyperparameters, e.g. also appearing in the mass distributions,  but these are omitted from the plot for the sake of visual clarity. Since the merger rate amplitude $\mathcal{R}_{\rm ref}$ parameterizes the merger rate at $z=0.2$ and $m_1=20M_\odot$, it will be almost solely determined by LIGO-Virgo's individual binary black hole detections, which happen at low to moderate redshifts. This is indeed the case, as illustrated by the recovery of the $\mathcal{R}_{\rm ref}$ parameter in Figure~\ref{fig:posteriorO3}, which is dominated by individual binary black hole events. In addition, the slope of the time-delay distribution $\kappa$ is constrained to negative values. A positive value of $\kappa$ would give rise to a merger rate $\mathcal{R}(z)$ that decreases with $z$ at low redshifts, contradicting the currently observed merger rate dependence on the redshift. Current individual binary black hole detections together with the gravitational-wave background estimator are not able to constrain the metallicity threshold $Z^{\max}$, nor the minimum time-delay parameter $t_d^{\rm min}$.

\cite{Fishbach_2021} similarly constrained the conditions of compact binaries using direct detections of binary mergers from the first two LIGO observing runs (but without additional constraints from the gravitational-wave background). We do note, however, that they assume an alternate star formation rate given by \cite{Madau_2017}, and that only two parameters are varied at a time, contrarily to our case where the slope, the minimum time delay, and the metallicity threshold are varied simultaneously. Keeping these differences in mind, our results are consistent with the ones obtained by \cite{Fishbach_2021}. They constrained the slope of the time delay distribution to negative values at the 90\% confidence level, and are unable to constrain the value of the metallicity threshold for the values of the minimum time delay considered in this paper.

\begin{figure*}
    \centering
    \includegraphics[scale = .46]{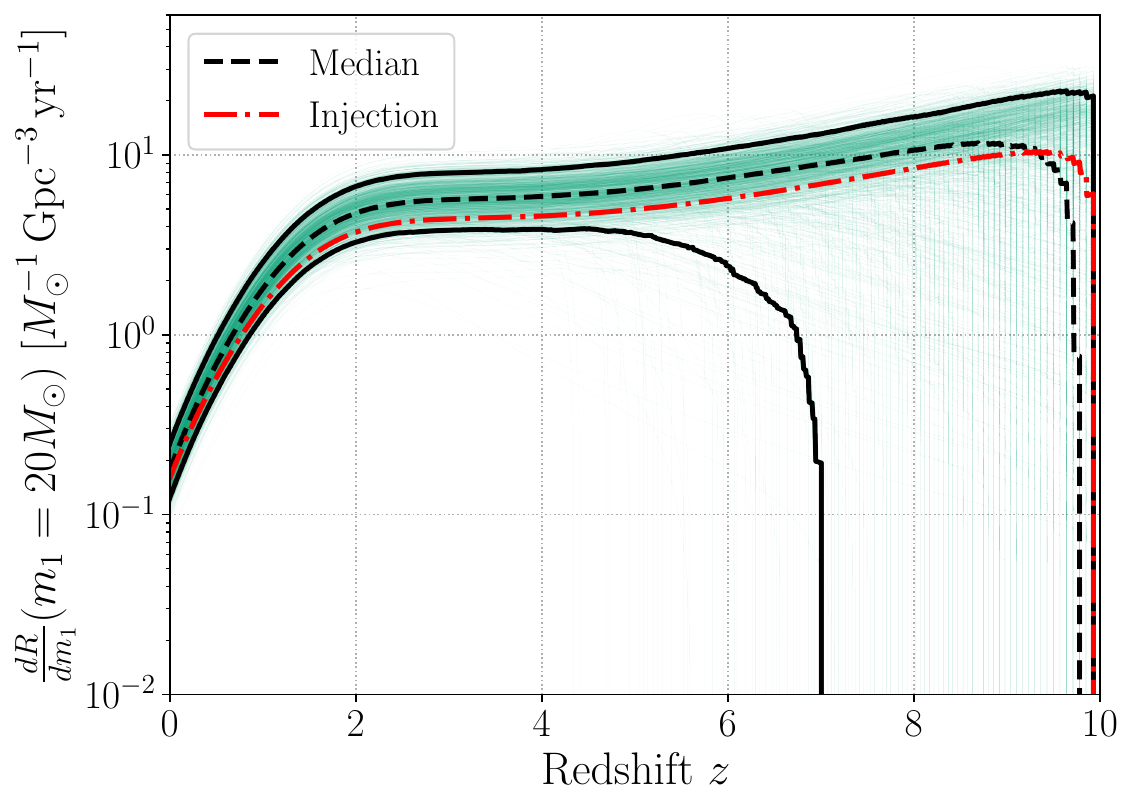}
    \includegraphics[scale = .46]{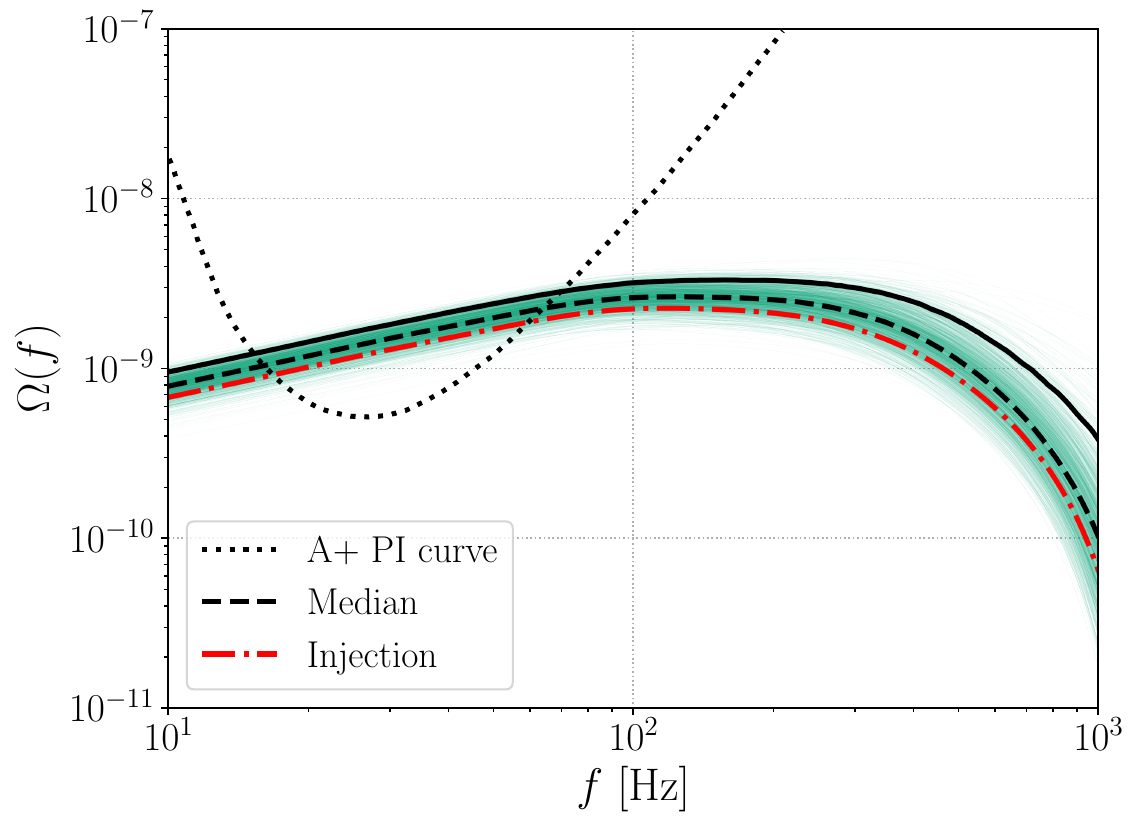}
    \caption{\textbf{Left}: Posterior samples for $\frac{dR}{dm_1}(m_1=20M_\odot)$ with the injected rate corresponding to a detectable gravitational-wave background at A+ sensitivity indicated by the red dash-dotted line. The full black lines denote the 5-95\% confidence interval, and the dashed black line denotes the median. \textbf{Right}: Posterior samples for $\Omega(f)$ with the injected detectable background denoted by the red dash-dotted line. The 95\% upper limit is denoted by the full black line, and the dashed black line represents the median. The dotted black line represents the $2\sigma$ A+ PI curve.}
    \label{fig:O5det}
\end{figure*}

In principle, measurements or upper limits on the gravitational-wave background are dominated by binaries at higher redshift, and could thus offer a complementary constraint on the slope of the time-delay distribution $\kappa$, the metallicity threshold $Z^{\max}$, and the minimum time delay $t_d^{\rm min}$. However, we find that current constraints on the gravitational-wave background are not sufficiently sensitive to appreciably change the measurements using solely individual binary black holes.
To illustrate this point further, we compare the resulting posterior samples on the merger rate $\frac{dR}{dm_1}(m_1=20M_\odot)$ in the left-hand side of Figure \ref{fig:mergerRateO3samples}. The individual blue traces correspond to individual posterior samples drawn from Figure \ref{fig:posteriorO3}, while solid black curves show the bounds obtained from analyzing binary black holes only. This shows that, at current sensitivity, results are entirely dominated by individual binary black hole mergers.

To illustrate why current upper limits on the gravitational-wave background are not yet informative, we show the expected $\Omega(f)$ from the posterior samples of the joint analysis using both direct detections and stochastic background constraints in the right-hand side of Figure \ref{fig:mergerRateO3samples}. We compare these samples with the $2\sigma$ O3 power-law integrated sensitivity curve (PI curve), which illustrates the sensitivity to an isotropic gravitational-wave background \citep{Thrane_2013}. Approximately, spectra lying above the PI curve are typically detectable with a signal-to-noise ratio $\ge1$, while spectra lying below have a signal-to-noise ratio $<1$. As can be seen, none of the posterior samples for $\Omega(f)$ are within reach of the PI curve, illustrating that we do not expect additional information from the gravitational-wave background contribution to the likelihood. 

Note that our results rely on our fairly strong systematic choice of a fixed star formation rate. As a check on the corresponding systematic uncertainty inherent in our results, we repeat this analysis in App.~\ref{app:resultsVangioni} for an alternative star formation rate given by \cite{10.1093/mnras/stu2600}, illustrating that our main conclusions remain unaltered when assuming a different star formation rate. The stellar metallicity distribution in Eq. \eqref{eq:metallicity_equation} represents an additional systematic uncertainty in our analysis. Although we use a fit provided by \cite{2006ApJ...638L..63L} in this work, more modern fits have recently been proposed \citep[e.g.,][]{2021MNRAS.508.4994C, van_Son_2023}. Future work will assess the impact of the choice of metallicity distribution on our results.



\subsection{Constraints using Advanced LIGO A+ sensitivity}
\label{ss:O5Results}

\begin{figure*}
    \centering
    \includegraphics[width = .9\textwidth]{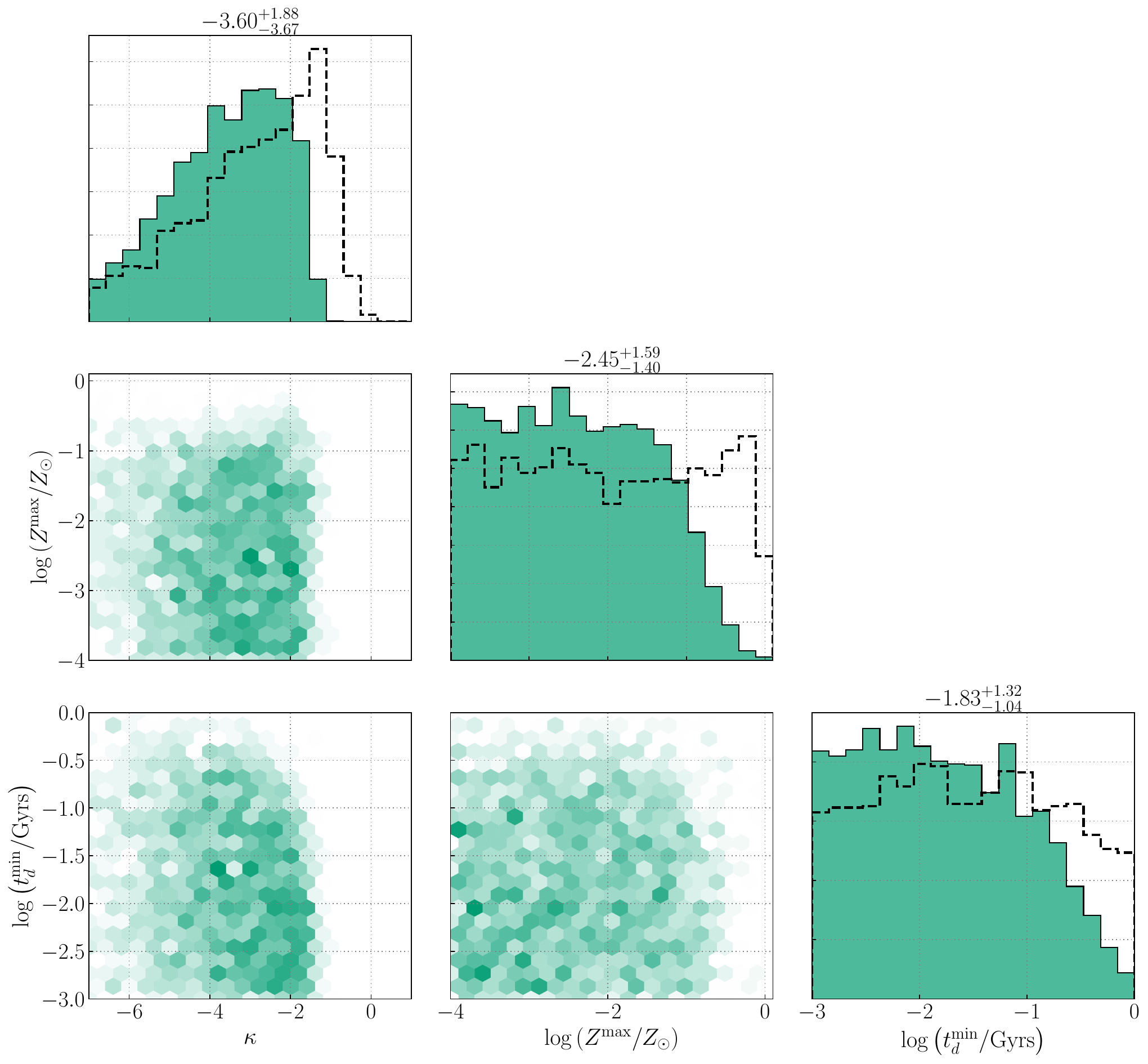}
    \caption{Posterior on the slope of the time-delay distribution $\kappa$, the metallicity threshold $Z^{\rm max}$, and the minimum time-delay parameter $t_d^{\rm min}$ for the joint analysis using both direct detections and stochastic background constraints at Advanced LIGO A+ sensitivity, in the presence of a \textit{detectable} gravitational-wave background. The dashed black lines represent the 1D histograms for the joint analysis on data from the first three observing runs from Figure~\ref{fig:posteriorO3}, for reference. We note that there is more support for larger negative values of the slope of the time-delay distribution $\kappa$, and smaller values of the metallicity parameter $Z^{\rm max}$ and the minimum time delay $t_d^{\rm min}$ are favored. In addition, we point out the complementarity of these constraints with the case of an undetectable gravitational-wave background at A+ sensitivity, as reported in Figure \ref{fig:posteriorO5_undet}.}
    \label{fig:posteriorO5_det}
\end{figure*}

\begin{figure*}
    \centering
    \includegraphics[scale = .46]{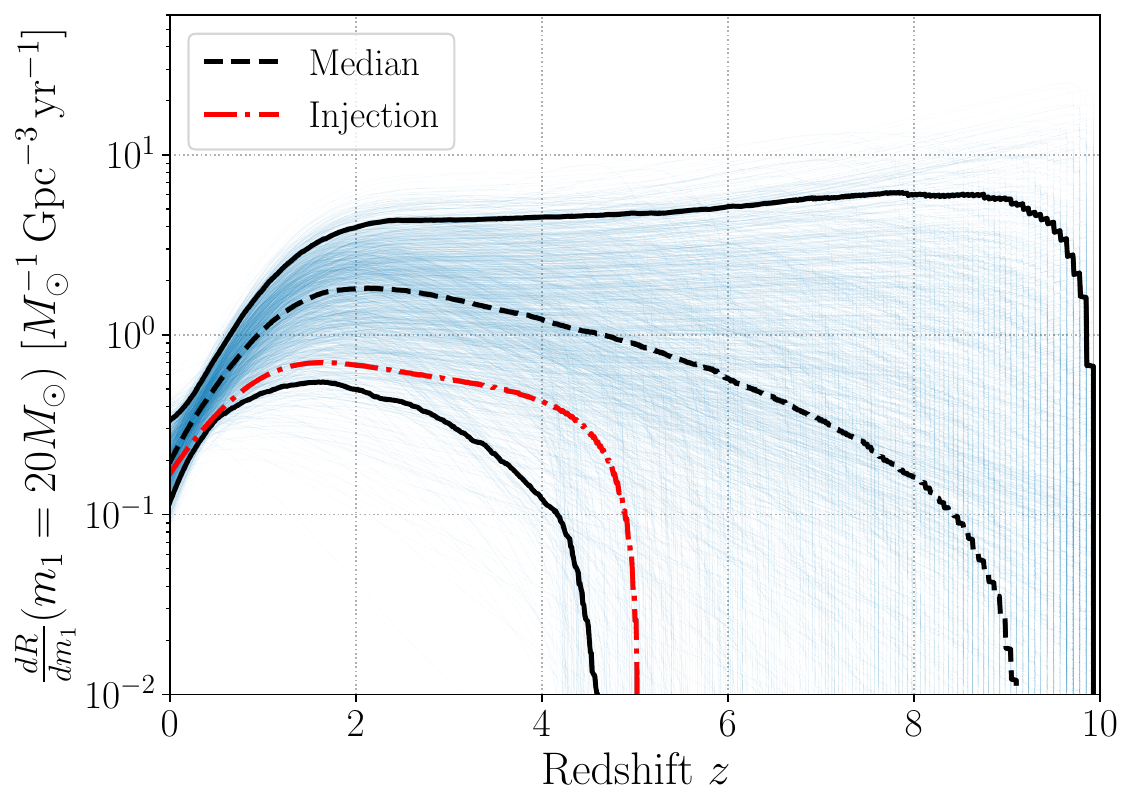}
    \includegraphics[scale = .46]{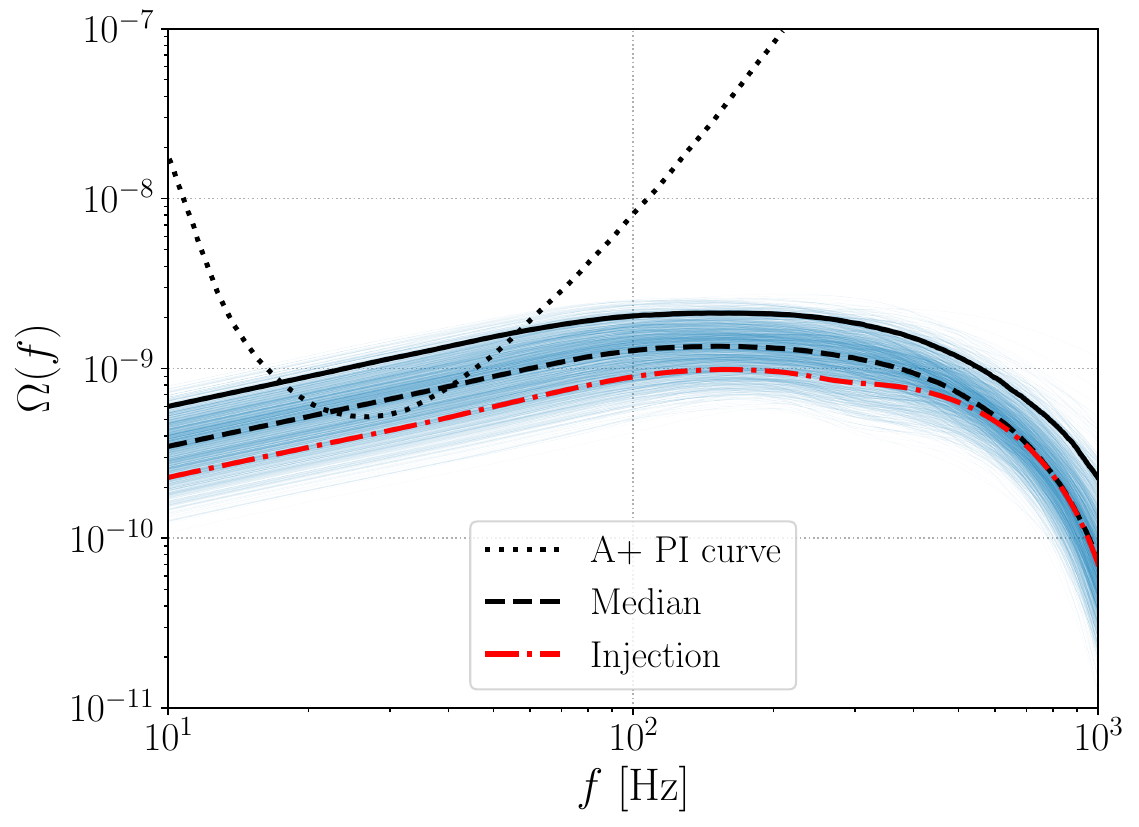}
    \caption{As in Figure~\ref{fig:O5det}, but now for the case of an \textit{undetectable} gravitational-wave background.}
    \label{fig:O5undet}
\end{figure*}

\begin{figure*}
    \centering
    \includegraphics[width = .9\textwidth]{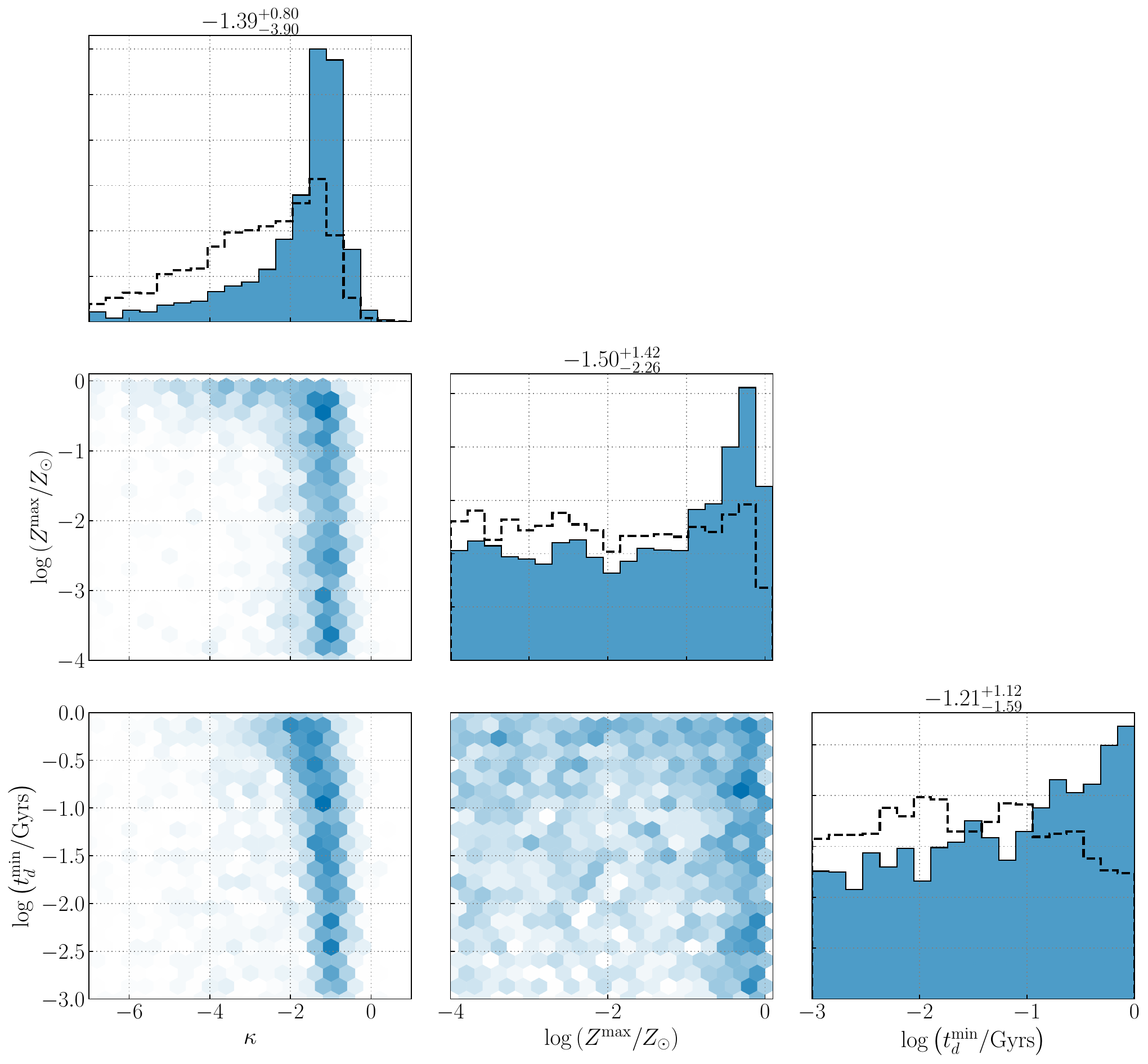}
    \caption{Posterior on the slope of the time-delay distribution $\kappa$, the metallicity threshold $Z^{\rm max}$, and the minimum time-delay parameter $t_d^{\rm min}$ for the joint analysis using both direct detections and stochastic background constraints at Advanced LIGO A+ sensitivity, in the presence of an \textit{undetectable} gravitational-wave background. The dashed black lines represent the 1D histograms for the joint analysis on data from the first three observing runs from Figure~\ref{fig:posteriorO3}, for reference. We note that large negative values for the slope of the time-delay distribution $\kappa$ are disfavored, and there is an enhanced support for larger values of the metallicity parameter $Z^{\rm max}$ and the minimum time delay $t_d^{\rm min}$. In addition, we point out the complementarity of these constraints with the case of a detectable gravitational-wave background at A+ sensitivity, as reported in Figure \ref{fig:posteriorO5_det}.}
    \label{fig:posteriorO5_undet}
\end{figure*}

Although current gravitational-wave background limits are not yet informative, increased integration time will give increasingly sensitive gravitational-wave background measurements, even in cases where detector horizons do not appreciably change. This increased sensitivity will be increasingly informative regarding the evolutionary history of binary black holes. We therefore consider stochastic background measurements obtainable with the Advanced LIGO detectors at their future A+ sensitivities \citep{AplusSensitivity}, assessing the potential of our joint analysis method in constraining the time-delay parameters. 

Following a future observing run with A+ LIGO instruments, there are two possibilities: a detection of the gravitational-wave background could be made, or the background could go undetected still. We will consider both cases, exploring the resulting constraints on the time-delay parameter space in each case. In both cases, we simulate a gravitational-wave background signal consistent with the results obtained in Sec.~\ref{ss:O3Results} using data from the first three observing runs. To accomplish this, we choose two samples from the posterior distribution presented in Figure \ref{fig:posteriorO3}: one posterior sample predicting a stochastic background that should be detectable with A+ LIGO, and a second predicting a background that should remain undetected. 
For both cases, we consider the LIGO Hanford and LIGO Livingston baseline, and assume an observation time of one year at the LIGO Advanced A+ sensitivity \citep{AplusSensitivity}. Future observing runs with more sensitive instruments will certainly also yield additional direct detections of binary black hole mergers. Our goal in this section, however, is primarily to understand the astrophysical information contained in a detection (or non-detection) of the stochastic background. We therefore do not simulate additional compact binary mergers, but use the same GWTC-3 catalog of direct detections as in Sec.~\ref{ss:O3Results}. 

{\textit{Case 1: A detection}} -- We first consider the detectable gravitational-wave background case, for which the simulated  $\Omega(f)$ spectrum is shown in Figure \ref{fig:O5det} (right), together with the $2\sigma$ A+ PI curve assuming a one-year observation time, for reference. As the injected gravitational-wave background lies above the PI curve, we indeed expect a detection. We infer the slope of the time-delay distribution $\kappa$, the metallicity threshold $Z^{\rm max}$, and minimum time delay $t_d^{\rm min}$ using both the individual binary black hole events from the GWTC-3 catalog, and the simulated detectable gravitational-wave background at A+ sensitivity. The posteriors for these parameters are shown in Figure \ref{fig:posteriorO5_det}, where the black dashed histograms represent the results of the analysis with current data reported in Figure \ref{fig:posteriorO3} to facilitate comparison. We omit the $\mathcal{R}_{\rm ref}$ posterior from the plot, and note that since the merger rate amplitude $\mathcal{R}_{\rm ref}$ is currently entirely determined by individual binary black hole events, the addition of gravitational-wave background measurement at A+ sensitivity does not alter the posterior on $\mathcal{R}_{\rm ref}$ compared to the runs at current sensitivity in the previous section. Contrarily, the posteriors on the time-delay parameters are much more constrained than in the runs using data from the first three observing runs of the previous section. Larger negative values are favored for the slope of the time-delay distribution $\kappa$, whereas smaller values are supported for both the metallicity parameter $Z^{\rm max}$, and the minimum time delay $t_d^{\rm min}$. In Figure \ref{fig:MergerRatePlot}, it was illustrated that smaller values of $Z^{\rm max}$, $\kappa$ and $t_d^{\rm min}$ corresponded to larger stochastic gravitational-wave background signals. A detection of a stochastic background by definition requires a reasonably large stochastic signal, pushing our posteriors on all three parameters to smaller values in Figure \ref{fig:posteriorO5_det}.

In addition to the posteriors, we show the merger rate $\frac{dR}{dm_1}(m_1=20M_\odot)$ constructed from the posterior samples in Figure \ref{fig:O5det} (left), as well as the posterior samples on $\Omega(f)$ in Figure \ref{fig:O5det} (right). The posterior samples of both quantities show agreement with the injected quantities, which are denoted by the red dash-dotted line in both figures.

{\textit{Case 2: A non-detection}} -- We now repeat the above analysis, but with a gravitational-wave background measurement at A+ sensitivity that falls below the expected sensitivity of a gravitational-wave background search, as displayed in Figure \ref{fig:O5undet} (right). This is illustrated by the fact that the injected gravitational-wave background falls below the A+ PI curve. Similarly to the detectable case, the simulated gravitational-wave background measurement is constructed from a posterior sample of the run with current data to ensure consistency with observed GWTC-3 binary black hole events. In Figure \ref{fig:posteriorO5_undet}, we show that the improved A+ sensitivity, even in the absence of a gravitational-wave background detection, allows for better constraints on the time-delay parameters. In particular, we note that the slope $\kappa$ of the time-delay distribution and the metallicity threshold $Z^{\rm max}$ are now required to lie along a preferred two-dimensional contour. The same is true of $\kappa$ and the minimum time delay $t_d^{\rm min}$. The result is that very large and negative values of $\kappa$, and small values of $Z^{\rm max}$ and $t_d^{\rm min}$ are excluded, 
since these values would all have yielded a detectable gravitational-wave background. Larger values of the metallicity parameter $Z^{\rm max}$ and the minimum time delay $t_d^{\rm min}$ are favored to be consistent with a non-detection of a gravitational-wave background. 
Furthermore, we note that $\frac{dR}{dm_1}(m_1=20M_\odot)$ is poorly constrained in the case a gravitational-wave background detection cannot be claimed, as illustrated in the left-hand side of Figure~\ref{fig:O5undet}. In particular, we point out the similarity of this posterior with the result with current data presented in the left-hand side of Figure~\ref{fig:mergerRateO3samples}, but note the somewhat lower 95\% upper limit due to the improved A+ sensitivity. Furthermore, the $\Omega(f)$ posterior in Figure \ref{fig:O5undet} (right) is consistent with the results using current data shown in the right-hand side of Figure \ref{fig:mergerRateO3samples}, although with a slightly lower 95\% upper limit, again as a result of the improved A+ sensitivity.

It is particularly interesting to stress that distinct regions of parameter space can be constrained depending on whether the gravitational-wave background is detected or not. The non-detection of a gravitational-wave background at A+ sensitivity would allow us to heavily constrain the slope of the time-delay distribution $\kappa$, as other values of that parameter would have given rise to a detectable gravitational-wave background, which was not observed. However, the $Z^{\rm max}$-$t_d^{\rm min}$ parameter space remains mostly unconstrained, although slightly favoring larger values of both parameters. In contrast, the detection of a gravitational-wave background instead allows us to place stronger bounds on $Z^{\rm max}$ and $t_d^{\rm min}$, but yields less stringent constraints on the $\kappa$ parameter, favoring larger negative values. 

\begin{figure}
    \hspace{-0.65cm}
    \includegraphics[scale=0.26]{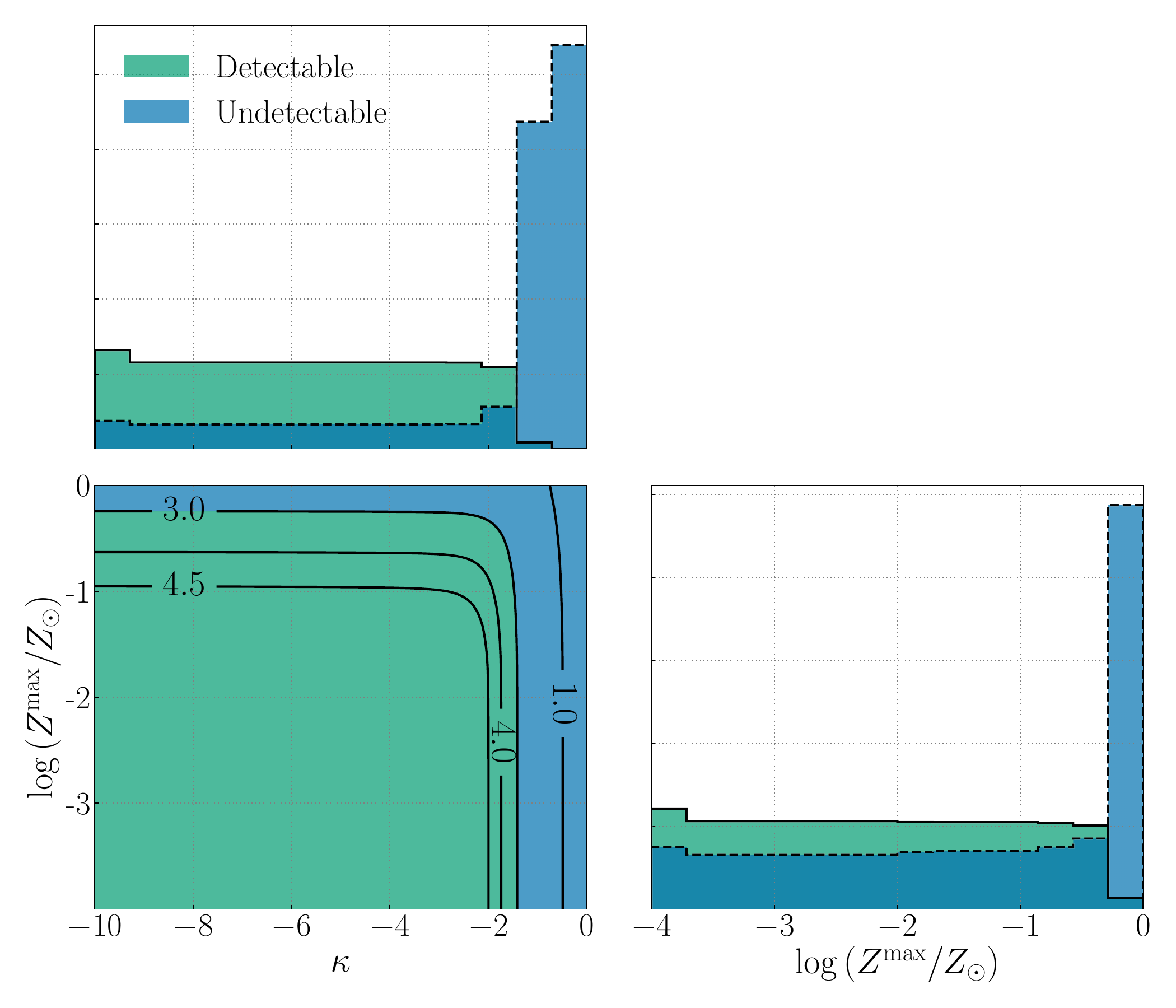}
    \caption{Contours of the expected stochastic signal-to-noise ratio using the Advanced A+ sensitivity for the slope of the time-delay distribution $\kappa$ and the maximum metallicity below which black holes are formed $Z^{\rm max}$. The 1D histograms denote the regions of parameter space where the signal-to-noise ratio is smaller (larger) than 3 in blue (green). In constructing the stochastic signal-to-noise ratio contours, the minimum time delay was set to $t_d^{\rm min}=0.05$ Gyrs and all other hyperparameters were set to the median value of the posteriors in Figure~\ref{fig:posteriorO3}. The regions of parameter space where the signal-to-noise ratio lies above (below) 3 corresponds to the region of parameter space where most posterior samples lie for a detectable (undetectable) gravitational-wave background in Figure~\ref{fig:O5det} (Figure~\ref{fig:O5undet}).}
    \label{fig:SNRplot}
\end{figure}

To illustrate why distinct regions of parameter space can be constrained with the detection or non-detection of a gravitational-wave background, we consider the signal-to-noise ratio given by
\begin{equation}
    \rho = \sqrt{2T}\frac{3H_0^2}{10\pi^2}\left(\int_{f_{\rm min}}^{f_{\rm max}}df~\frac{\gamma_{12}^2(f)\Omega^2(f)}{f^6P_{1}(f)P_{2}(f)}\right)^{1/2},
\end{equation}
where $T$ denotes the observation time, $\gamma_{12}$ the overlap reduction function, and $P_{i}$ the noise power spectral density of detector $i=\{1,~2\}$ \citep{Romano_2017}. In Figure~\ref{fig:SNRplot}, we show contours of the stochastic signal-to-noise ratio in the $\kappa-Z^{\rm max}$ parameter space, using the Advanced LIGO A+ sensitivity. When computing the expected gravitational-wave background spectrum $\Omega(f)$, the minimum time delay was set to $t_d^{\rm min}=0.05$ Gyrs and the values of other hyperparameters were set to the median values of the posteriors in Figure~\ref{fig:posteriorO3}. Additionally, we show 1D histograms of regions of parameter space where the expected signal-to-noise ratio is smaller (larger) than 3 in blue (green). In particular, we want to highlight that the regions of parameter space where the signal-to-noise ratio is larger (smaller) than 3 corresponds to the region of parameter space where most posterior samples lie for a detectable (undetectable) gravitational-wave background in Figure~\ref{fig:O5det} (Figure~\ref{fig:O5undet}). The detection (non-detection) effectively forces the posterior samples to lie within a region of parameter space with a corresponding signal-to-noise ratio that is above (below) the detection threshold of $\rho=3$.

Therefore, the two cases considered in this work, both a gravitational-wave background detection and non-detection at the Advanced LIGO A+ sensitivity, provide complementary information about the time-delay distribution and metallicity-specific star formation rate.

\section{Conclusion}
\label{s:Conclusion}
The target of this work is to shed light on the evolution of binary black holes with gravitational-wave data. We currently have two sources of information -- the direct binary black hole detections comprised in GWTC-3, and the upper limits on the stochastic gravitational-wave background. Each of these observational inputs provides us information about the binary merger rate at different redshifts. Our goal is to synthesize both together to constrain the parameters that govern the metallicity-specific formation rate of binary black hole progenitors and their subsequent evolutionary time delays


We consider LIGO-Virgo data from the first three observing runs, including both individual binary black hole mergers from the GWTC-3 catalog, as well as current upper limits on the gravitational-wave background. Current individual binary black hole mergers allow to constrain the slope of the time-delay distribution to negative values, but leave the other parameters of interest unconstrained. The addition of information from the gravitational-wave background in our joint analysis using individual binary black hole mergers and the upper limits on a gravitational-wave background does not provide more stringent constraints at current sensitivity.

Nevertheless, we consider the future Advanced LIGO A+ sensitivity and consider both the case of a gravitational-wave background detection and non-detection at that sensitivity. We show that both cases offer unique and complementary constraints on the parameter space of interest. Indeed, the non-detection of a gravitational-wave background at Advanced LIGO A+ sensitivity results in tight constraints on the slope of the time-delay distribution. On the other hand, the detection of a gravitational-wave background would allow to constrain larger parameter space regions of the metallicity threshold and minimum time-delay parameter.

Although considering current gravitational-wave background upper limits does not improve the constraints on the metallicity-specific star formation rate and time-delay parameters at current sensitivity, this work shows that the additional information contained in a gravitational-wave background measurement will be essential to learn more about the environment in which compact binaries formed and their evolution.

\section*{Acknowledgement}
Kevin Turbang is supported by FWO-Vlaanderen through grant number 1179522N. We thank Martyna Chruślińska for useful discussions. We are grateful for the useful comments given by Maya Fishbach, Shanika Galaudage, and Christopher Berry. This research has made use of data, software and/or web tools obtained from the Gravitational Wave Open Science Center
(https://www.gw-openscience.org), a service of LIGO
Laboratory, the LIGO Scientific Collaboration and the
Virgo Collaboration. Virgo is funded by the French
Centre National de Recherche Scientifique (CNRS), the
Italian Istituto Nazionale della Fisica Nucleare (INFN)
and the Dutch Nikhef, with contributions by Polish and
Hungarian institutes. This material is based upon work
supported by NSF’s LIGO Laboratory which is a major
facility fully funded by the National Science Foundation. The authors are grateful for computational resources provided by the LIGO Laboratory and supported by NSF Grants PHY-0757058 and PHY-0823459.

\textit{Data and code:} Data produced in this paper are available at \url{https://zenodo.org/doi/10.5281/zenodo.10016289}. The code used to produce the results in this paper can be found on \url{https://github.com/kevinturbang/bbh_gwb_time_delay_inference/tree/main}.

\textit{Software:} {\tt jax} \citep{jax2018github}, {\tt NumPyro} \citep{phan2019composable}, {\tt matplotlib} \citep{Hunter:2007}, {\tt scipy} \citep{2020SciPy-NMeth}, {\tt numpy} \citep{harris2020array}, {\tt h5py} \citep{andrew_collette_2021_4584676}

\appendix
\section{Results with an alternative star formation history}
\label{app:resultsVangioni}

In order to turn gravitational-wave observations into constraints on evolutionary conditions (their metallicity dependence and time delays) of binary black holes, it was necessary to \textit{assume} an underlying global star formation rate. In the main text, we used the \cite{MDSFR} model, but the global star formation rate remains highly uncertain, and thus represents a source of systematic uncertainty in our analysis. In this section, we explore this systematic uncertainty by repeating our analyses under a different model for the global star formation rate. We find that our conclusions in the main text remain robust under such alternative assumptions.

\begin{figure}[h!]
    \includegraphics[scale=0.47]{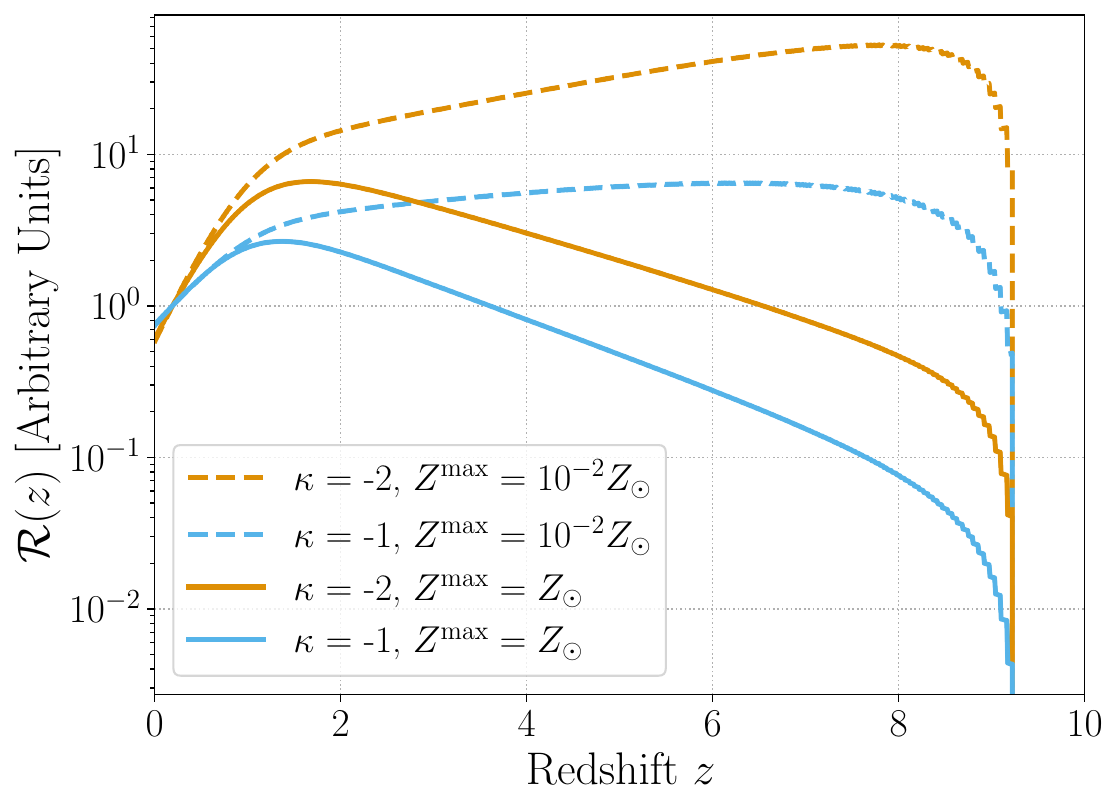}
    \includegraphics[scale=0.47]{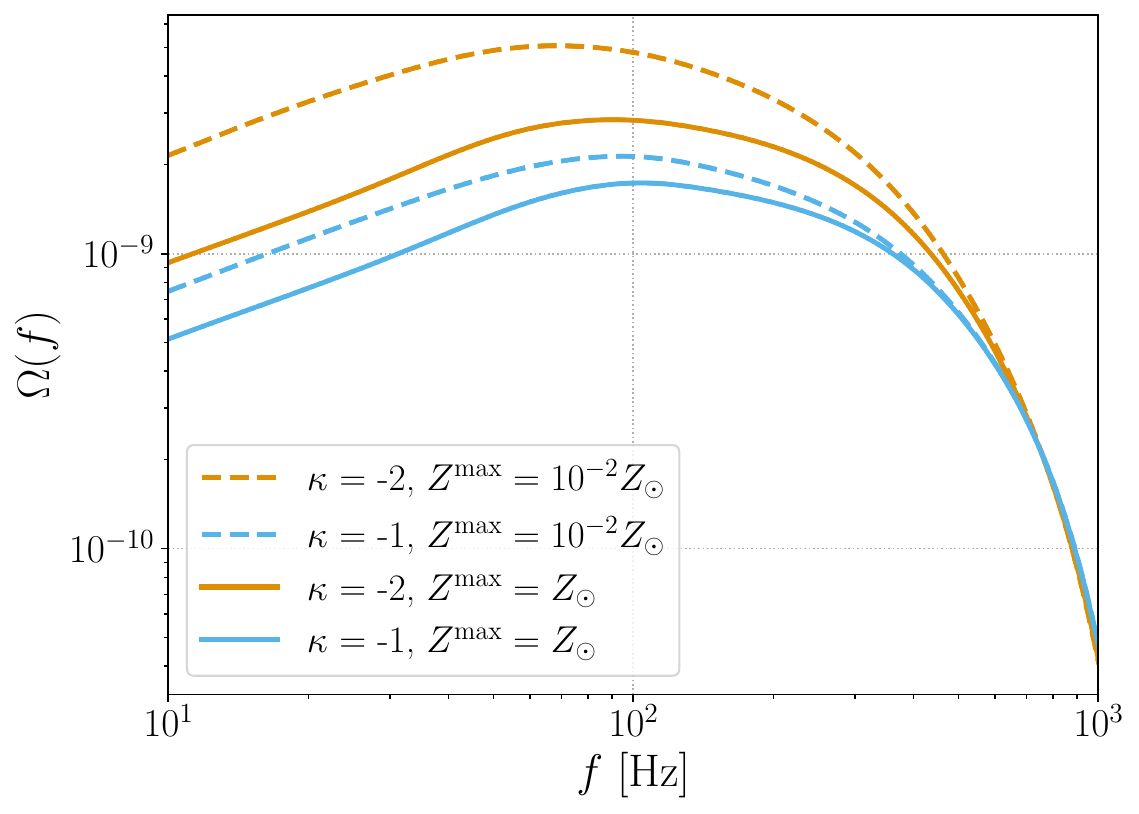}
    \caption{As in Figure \ref{fig:MergerRatePlot}, but using an alternative global star formation rate given by \cite{10.1093/mnras/stu2600}.}
    \label{fig:MergerRatePlotVA}
\end{figure}

We consider a global star formation rate as given by \cite{10.1093/mnras/stu2600}:
\begin{equation}
\label{eq:SFR-Vangioni}
    R_*(z)\propto \frac{\exp(b(z-z_0))}{a-b+b\exp(a(z-z_0))},
\end{equation}
where $a=2.8$, $b=2.46$, and $z_0=1.72$. We show an example of the corresponding merger rate $\mathcal{R}(z)$ and the expected $\Omega(f)$ spectrum in Figure \ref{fig:MergerRatePlotVA}, in analogy to Figure \ref{fig:MergerRatePlot}. As the parameters of the underlying distributions are varied, both the merger rate $\mathcal{R}(z)$ and the $\Omega(f)$ spectrum vary similarly to the ones discussed in the main text. We therefore refrain from providing a detailed discussion here about the various parameters, and refer to the main text for more details.

Using this alternate star formation rate, we repeat our inference of the binary black hole metallicity dependence and time-delay distribution, using both black hole detections in GWTC-3 and the latest constraints on the stochastic background. The results are effectively identical to those shown previously in Figure \ref{fig:posteriorO3} using the \cite{MDSFR} model. We therefore do not show them here. Instead, we will look ahead to constraints possible with future A+ LIGO instruments, as in Sec. \ref{ss:O5Results}.

\begin{figure*}[h!]
    \centering
    \includegraphics[width = .9\textwidth]{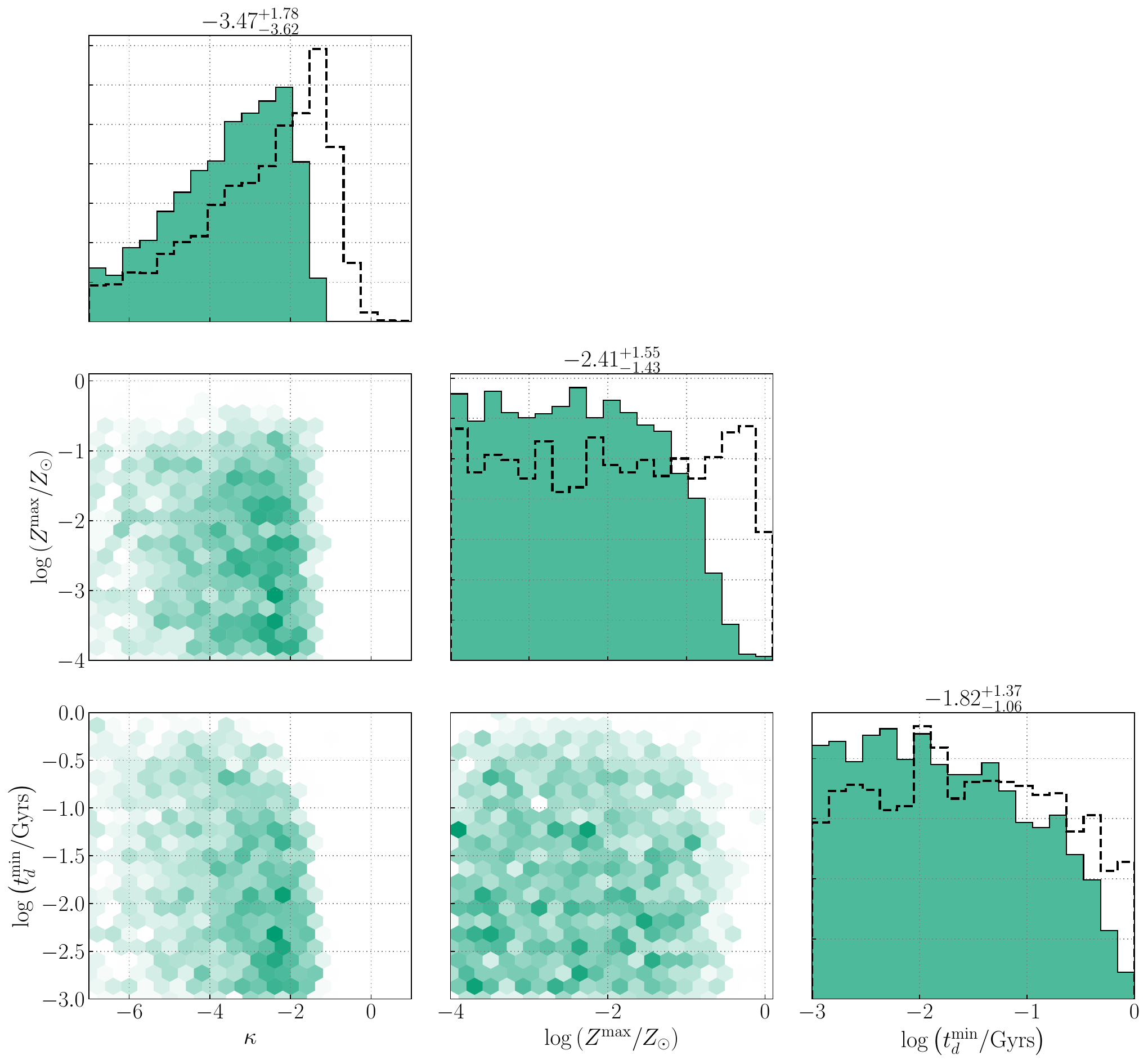}
    \caption{Posterior on the slope of the time-delay distribution $\kappa$, the metallicity threshold $Z^{\rm max}$, and the minimum time-delay parameter $t_d^{\rm min}$ for the joint analysis using both direct detections and stochastic background constraints at Advanced LIGO A+ sensitivity, in the presence of a detectable gravitational-wave background. The dashed black lines represent the 1D histograms for the joint analysis on data from the first three observing runs, for reference. We note that there is more support for larger negative values of the slope of the time-delay distribution $\kappa$, and smaller values of the metallicity parameter $Z^{\rm max}$ and the minimum time delay $t_d^{\rm min}$ are favored. In addition, we point out the complementarity of these constraints with the case of an undetectable gravitational-wave background at A+ sensitivity, as reported in Figure \ref{fig:posteriorO5VA_undet}.}
    \label{fig:posteriorO5VA_det}
\end{figure*}

\begin{figure*}[h!]
    \centering
    \includegraphics[width = .9\textwidth]{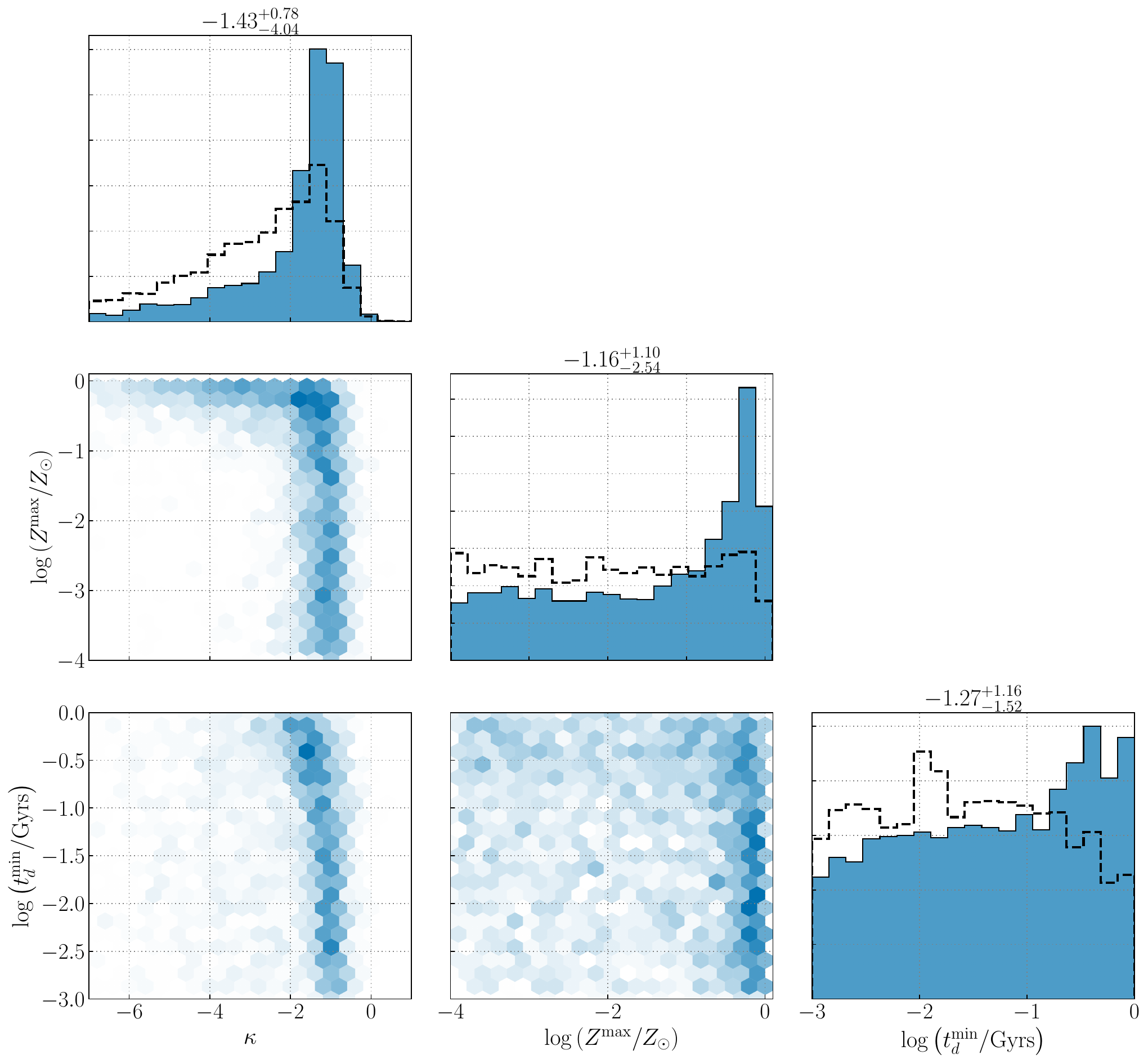}
    \caption{Posterior on the slope of the time-delay distribution $\kappa$, the metallicity threshold $Z^{\rm max}$, and the minimum time-delay parameter $t_d^{\rm min}$ for the joint analysis using both direct detections and stochastic background constraints at Advanced LIGO A+ sensitivity, in the presence of an undetectable gravitational-wave background. The dashed black lines represent the 1D histograms for the joint analysis on data from the first three observing runs, for reference. We note that large negative values for the slope of the time-delay distribution $\kappa$ are disfavored, and the support at larger values of the metallicity parameter $Z^{\rm max}$ and the minimum time delay $t_d^{\rm min}$. In addition, we point out the complementarity of these constraints with the case of an detectable gravitational-wave background at A+ sensitivity, as reported in Figure \ref{fig:posteriorO5VA_det}.}
    \label{fig:posteriorO5VA_undet}
\end{figure*}

Similarly to the analysis in the main text, we report the posteriors for the joint analysis using both direct detections and stochastic background constraints using the \cite{10.1093/mnras/stu2600} star formation rate for a simulated gravitational-wave background at A+ sensitivity, both detectable and undetectable. The analogue of Figures \ref{fig:posteriorO5_det} and \ref{fig:posteriorO5_undet} in the main text are reported in Figures \ref{fig:posteriorO5VA_det} and \ref{fig:posteriorO5VA_undet}. Comparing these sets of figures, we see that our qualitative results show little dependence on the exact model used for global star formation. Once more, we see that an undetectable gravitational-wave background at A+ sensitivity results in strong constraints on the slope of the time-delay distribution, but leaves the metallicity and the minimum time-delay parameters unconstrained. Conversely, a detectable gravitational-wave background at A+ sensitivity still results in larger regions of the metallicity and minimum time-delay parameter space being excluded, and shows support for larger negative values of the slope of the time-delay distribution. Therefore, also with the \cite{10.1093/mnras/stu2600} star formation rate, both cases at future Advanced LIGO A+ sensitivity would yield distinct and complementary information about the environment in which binaries formed and their evolution throughout time.

In addition, our choice of model for the cumulative metallicity of star formation, as given by Eq. \eqref{eq:metallicity_equation}, represents a similar systematic source of uncertainty. Future work will explore how variations in metallicity models similarly impact our conclusions.

\section{Additional distributions and priors}
\label{app:models}
In addition to the mass and redshift distribution described in the main text, we provide the assumed distributions for the spin magnitudes and spin-orbit tilt angles. These are modeled as truncated Gaussian distributions \citep{callister2023parameterfree}:
\begin{equation}
    \pi(\chi_i)=\sqrt{\frac{2}{\pi\sigma_\chi^2}}\frac{e^{-(\chi_i-\mu_\chi)^2/2\sigma_\chi^2}}{{\rm Erf}\left(\frac{1-\mu_\chi}{\sqrt{2\sigma_\chi^2}}\right)+{\rm Erf}\left(\frac{\mu_\chi}{\sqrt{2\sigma_\chi^2}}\right)},
\end{equation}
where $\mu_\chi$ stands for the mean of the distribution, $\sigma_\chi^2$ for the variance, and
\begin{equation}
    \pi(\cos\theta_i)=\sqrt{\frac{2}{\pi\sigma_u^2}}\frac{e^{-(\cos\theta_i-1)^2/2\sigma_u^2}}{{\rm Erf}\left(\frac{-2}{\sqrt{2\sigma_u^2}}\right)},
\end{equation}
where the mean is assumed to be 1, and the variance $\sigma_u^2$ is inferred from the data.

Furthermore, we provide a list of priors used for the parameter estimation performed in this work in Table \ref{tab:priors}.
\begin{table}[h!]
    \centering
    \begin{tabular}{c|c|c|c}
    
    \hline
    \hline
    \multicolumn{4}{c}{\textbf{Mass distribution}}\\
    \hline
    \hline
    Parameter & Prior & Minimum & Maximum\\
    \hline
        $m_{\rm low}/M_\odot$ & Uniform & 5 & 15\\ 
        $m_{\rm high}/M_\odot$ & Uniform & 50 & 100\\
        $\mu_{m}/M_\odot$ & Uniform & 20 & 50 \\
        $\sigma_{m}/M_\odot$ & Uniform & 1.5 & 15\\
        $f_{\rm peak}$ & Log-Uniform & $10^{-3}$&1\\
        $\delta m_{\rm low}/M_\odot$ & Log-Uniform & $10^{-1}$&$10^{0.5}$\\
        $\delta m_{\rm high}/M_\odot$ & Log-Uniform & $10^{0.5}$&$10^{1.5}$\\
        
    \hline
    \hline
    Parameter & Prior & Mean & Standard deviation\\
    \hline
        $\alpha$ & Gaussian & -2 & 3\\
        $\beta_q$ & Gaussian & 0 & 3\\

    \hline
    \hline
    \multicolumn{4}{c}{\textbf{Time-delay distribution}}\\
    \hline
    \hline
    Parameter & Prior & Minimum & Maximum\\
    \hline
    $\mathcal{R}_{\rm ref}/M_\odot^{-1}{\rm Gpc^{-3}yr^{-1}}$ & Log-Uniform & $10^{-2}$ & $10$\\
    $Z^{\rm max}/Z_\odot$ & Log-Uniform & $10^{-4}$ & 1\\
    $t_d^{\rm min}$/Gyrs & Log-Uniform & $10^{-3}$&1\\
    \hline
    \hline
    Parameter & Prior & Mean & Standard deviation\\
    \hline
    $\kappa$ & Gaussian & -1 & 3\\
    \hline\hline
    \multicolumn{4}{c}{\textbf{Spin distribution}}\\
    \hline
    \hline
    Parameter & Prior & Minimum & Maximum\\
    \hline
    $\mu_\chi$ & Uniform & 0 & 1\\
    $\sigma_\chi$ & Log-Uniform & $10^{-1}$ & 1\\
    $\sigma_u$ & Uniform & 0.3 & 2\\
    \hline
        
    \hline\hline

    \end{tabular}
    \caption{Prior choice for the hyperparameters describing the mass distributions, the time-delay distribution and the spin distribution. Note that, for the minimum time-delay parameter $t_d^{\rm min}$, we do not extend the prior to lower values to not be in conflict with the minimum lifetime of massive stars \citep[e.g.,][]{2013sse..book.....K}.}
    \label{tab:priors}
\end{table}

\section{Data}
\label{app:Data}
In this appendix, we provide more detailed information regarding the exact inputs to our analysis. In our analyses, we include the binary black holes of the GWTC-3 catalog \citep{PhysRevX.13.011048} detected with a false-alarm rate below 1 yr$^{-1}$. In the GWTC-3 catalog, there are two events, GW190814 \citep{Abbott_2020} and GW190917, that are possibly binary
black holes but which are known to be outliers with respect to the bulk binary population \citep{PhysRevX.13.011048}. We do not consider these two events, leaving 69 binary black holes to be included in our analysis. Publicly-available parameter estimation samples are used, which are provided by the Gravitational-Wave Open Science Center\footnote{https://www.gw-openscience.org/} \citep{Vallisneri_2015,Rich_Abbott_2021,KAGRA:2023pio} and/or Zenodo. We use the ``Overall posterior'' samples\footnote{https://dcc.ligo.org/LIGO-P1800370/public} for GWTC-1 events \citep{PhysRevX.9.031040}, use the ``PrecessingSpinIMRHM'' samples for events from GWTC-2 \citep{PhysRevX.11.021053}\footnote{https://dcc.ligo.org/LIGO-P2000223/public}, and use use the ``C01:Mixed'' samples\footnote{https://zenodo.org/record/5546663} for new events in GWTC-3 \citep{PhysRevX.13.011048}. These contain the combination of parameter estimation samples from several waveform families, each of which include the physical effects of spin precession. We note that the GWTC-2 and GWTC-3 samples additionally include higher-order mode content, whereas GWTC-1 samples, in contrast, do not.

In our analysis, injected signals are used to characterize selection effects, as given by Eq.~\eqref{eq:MC_selection_effects}. We use the injection set\footnote{https://zenodo.org/record/5636816} detailed in \cite{PhysRevX.13.011048}, labeling injections as ``found'' when they are recovered with false-alarm rates below 1 yr$^{-1}$ in at least one search pipeline. However, the injections for the O1 and O2 observing runs do not have associated false-alarm rates, only network signal-to-noise ratios $\rho$. We therefore consider them ``found'' if $\rho\ge10$.

In addition, for the joint analysis using both direct detections and stochastic background constraints at current O3 sensitivity, we use the results of the search for an isotropic gravitational-wave background following the LIGO-Virgo O3 observing run\footnote{https://dcc.ligo.org/LIGO-G2001287/public} \citep{Abbott_2021}. Note that we only consider the LIGO Hanford -- LIGO Livingston baseline, since the LIGO-Virgo baselines have negligible sensitivity to CBC signals.

\section{Computation of $\Omega(\lowercase{f})$}
\label{app:computationOmega}
The computation of the expected $\Omega(f)$, given a set of hyperparameters $\Lambda$, can be written as in Eq. \eqref{eq:OmegaBBH} in the main text. We recall that this depends on the average energy $\left\langle\frac{d E_{s}}{d f_{s}}\right\rangle$ radiated by each binary, given by Eq. \eqref{eq:AverageEnergy}, where the integration is performed over the masses $m_1$ and $m_2$. Therefore, including the integration over redshift, the computation of the expected $\Omega(f)$ requires the evaluation of a three-dimensional integral. Numerically, the integral can be evaluated on a grid, resulting in a fairly inefficient and slow implementation of the computation. Alternatively, one can rely on a Monte Carlo integration technique to evaluate the integral, where one relies on averaging many individual draws of the population. 

In our implementation of this approach, $N$ samples are drawn from a uniform distribution at the start of the computation for each of the parameters $z$, $m_1$, and $m_2$. We denote the i-th parameter draw as $\phi^i=\{z^i,m_1^i,m_2^i\}$. We proceed by computing the energy spectra $dE_s/df_s$ for each of these parameter draws and label these $dE^i_s/df_s$. However, we want to compute the $\Omega(f)$ for the population corresponding to the distributions described in the main text, with hyperparameters $\Lambda$. We therefore rely on reweighting $dE^i_s/df_s$ given some hyperparameters $\Lambda$. Schematically, the computation takes the form
\begin{equation}
    \label{eq:MCOmega}
    \Omega(f) \sim \frac{1}{N}\sum_i^{N} w_i \frac{dE^i_s}{df_s},
\end{equation}
where the average is taken over the number of sample draws $N$, and the weights $w_i$ are defined as
\begin{equation}
    w_i = \frac{\mathcal{R}(z^i)(1+z^i)^{-1}H(z^i)^{-1}}{p_{\rm draw}(z^i)}\frac{\phi(m_1^i)}{p_{\rm draw}(m_1^i)}\frac{p(m_2^i)}{p_{\rm draw}(m_2^i)}.
\end{equation}
The $p_{\rm draw}$ distributions represent the uniform distributions used to make the initial draws of the parameters $\phi^i$, whereas the numerators denote the distributions we are reweighting to, characterizing the population, corresponding to a set of hyperparameters $\Lambda$, for which we want to compute the $\Omega(f)$ spectrum.

\begin{figure}[h!]
    \hspace{-1cm}
    \includegraphics[scale=.335]{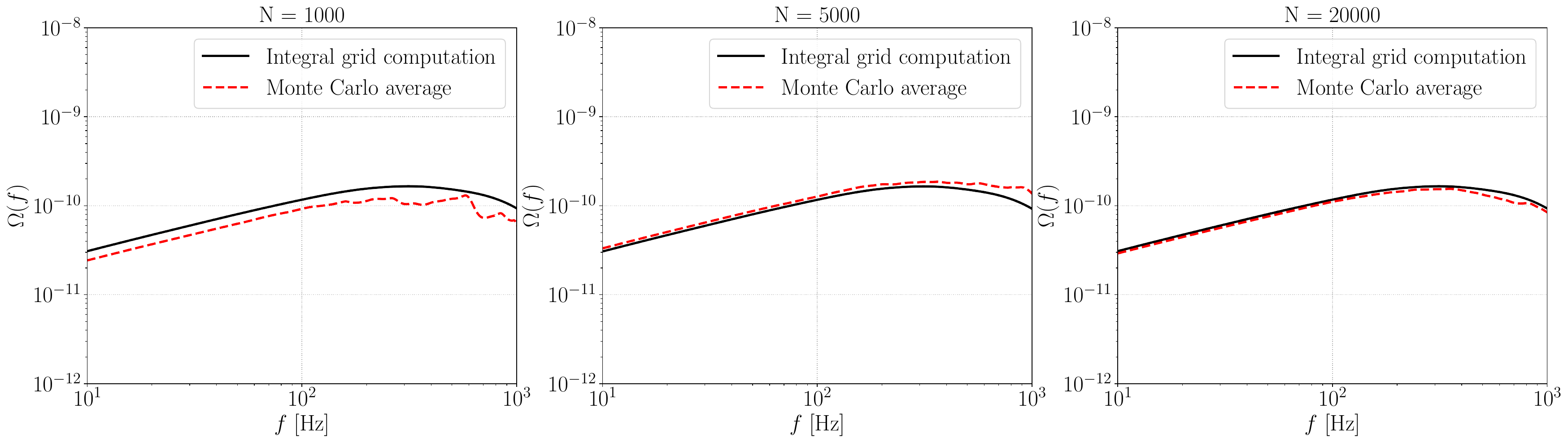}
    \caption{Convergence of the $\Omega(f)$ spectrum using the Monte Carlo average approach, compared to the computation evaluating the integrals in Eq.~\eqref{eq:OmegaBBH} directly. The black line denotes the $\Omega(f)$ spectrum computed using by evaluating the integrals numerically on a grid, whereas the red dashed lines show the result of the Monte Carlo approach for different numbers of sample draws. For illustrative purposes, several numbers of sample draws $N$ are used in the Monte Carlo average, namely 1000, 5000, and 20000. It was found that $N=20000$ approximates the $\Omega(f)$ spectrum well enough.}
    \label{fig:MCOmegaComputation}
\end{figure}

In order to show the convergence of the Monte Carlo average, we show the $\Omega(f)$ spectrum computed by performing the integral explicitly and compare it to our implementation for several values of the number of initial draws. The results are displayed in Figure~\ref{fig:MCOmegaComputation}. Note that although the results are only shown for one set of hyperparameters, i.e., one specific distribution, and therefore, one $\Omega(f)$ spectrum in the plot, tests were performed on various other distributions to verify convergence. In general, a value $N=20000$ approximates the full integral computation well in the frequency range of interest for gravitational-wave background searches with the LIGO-Virgo-KAGRA detectors, with relative errors of $\mathcal{O}(10^{-1}-10^{-2})$. Therefore, throughout our work, we use $N=20000$ when computing the $\Omega(f)$ spectrum with the Monte Carlo averaging procedure.

\newpage
\bibliography{sample631}{}
\bibliographystyle{aasjournal}

\end{document}